\documentclass[twocolumn]{aastex63}

\usepackage{graphicx}
\usepackage{dcolumn}
\usepackage{bm}
\usepackage[varvw]{newtxmath}
\usepackage{ulem}

\begin{document}

\shorttitle{The Galactic Center Reference Frame}
\shortauthors{Darling et al.}

\title{The Galactic Center Reference Frame:  A 30-Year Solution}

\author[0000-0003-2511-2060]{Jeremy Darling}
 \affiliation{Center for Astrophysics and Space Astronomy,
Department of Astrophysical and Planetary Sciences\\
University of Colorado, 389 UCB,
Boulder, CO 80309-0389, USA}
 \email{jeremy.darling@colorado.edu}

 \author[0000-0002-7517-9223]{Jennie Paine}
\affiliation{Center for Space Sciences and Technology, University of Maryland, Baltimore County, Baltimore, MD 21250, USA}
\affiliation{Observational Cosmology Lab, NASA Goddard Space Flight Center, Greenbelt, MD 20771, USA}
\affiliation{Center for Research and Exploration in Space Science and Technology, NASA/GSFC, Greenbelt, MD 20771, USA}

\author[0000-0003-2874-1196]{Matthew Hosek, Jr.}
\altaffiliation{Brinson Prize Fellow}
\affiliation{UCLA Department of Physics and Astronomy, Los Angeles, CA 90095, USA}

\author[0000-0001-5972-663X]{Shoko Sakai}
\affiliation{UCLA Department of Physics and Astronomy, Los Angeles, CA 90095, USA}

\author[0000-0002-5708-0481]{Stefan Gillessen}
\affiliation{Max Planck Institute for Extraterrestrial Physics, Giessenbachstra\ss e 1, 85748 Garching, Germany}

\author[0000-0002-0647-718X]{Tatsuya Akiba-Lewis}
\affiliation{UCLA Department of Physics and Astronomy, Los Angeles, CA 90095, USA}

\author[0000-0003-3230-5055]{Andrea Ghez}
\affiliation{UCLA Department of Physics and Astronomy, Los Angeles, CA 90095, USA}

\author{Frank Eisenhauer}
\affiliation{Max Planck Institute for Extraterrestrial Physics, Giessenbachstra\ss e 1, 85748 Garching, Germany}
\affiliation{Department of Physics, TUM School of Natural Sciences, Technical University of Munich, 85747 Garching, Germany}

\author[0000-0002-2767-9653]{Reinhard Genzel}
\affiliation{Max Planck Institute for Extraterrestrial Physics, Giessenbachstra\ss e 1, 85748 Garching, Germany}
\affiliation{Departments of Physics \& Astronomy, Le Conte Hall, University of California, Berkeley, CA 94720, USA}

 \begin{abstract}
   Stellar orbits probe the metric of and mass around the Sgr A* massive black hole.
   To enable study of the black hole, the surrounding mass, and tests of gravity, a precise reference
   frame must be established and maintained. 
   We present an updated stellar maser-based reference frame for the inner parsec of the Galactic Center, referenced to
   the radio continuum of Sgr A*.
   This work synthesizes 30 years of astrometric observations of 15 SiO stellar masers and the Sgr A* continuum at 43~GHz.
  The uncertainty in the mean offset of the maser ensemble from Sgr A* is 0.08~mas in the current epoch and will decay to 0.10~mas by 2030.
   At a distance of 8.2 kpc, these correspond to position uncertainties of 0.6 AU and 0.8 AU, respectively, which is similar to or smaller than the stellar SiO-emitting region.
   The mean maser proper motion uncertainties are 0.05 mas yr$^{-1}$ in RA and 0.08 mas yr$^{-1}$ in Dec, corresponding to 
   1.8 km s$^{-1}$ and 3.2 km s$^{-1}$.
   The reference frame stability between IR and maser astrometry
   is 10~$\mu$as~yr$^{-1}$ (0.40 km s$^{-1}$) in RA and 15 $\mu$as~yr$^{-1}$ (0.57 km s$^{-1}$) in Dec.
   The masers show substantial variability in flux density and velocity structure, most notably IRS~7, which shows a 30-fold brightening in a blueshifted component since 2022.
   This variability may set a floor on future astrometry improvements and especially measurements of radial acceleration.
\end{abstract}

\section{\label{sec:intro}Introduction}

Near-infrared observations of the orbits of stars close to Sgr A* conclusively demonstrated the presence of a massive black hole in the center
of the Galaxy \citep{ghez2008,genzel2010}.
This work relied on an astrometric reference frame enabled by simultaneous radio observations of stellar masers and the Sgr A* radio
continuum \citep{menten1997,reid2003,reid2007}.  Since the black hole was not historically visible in the near-IR imaging, measuring the
offsets between IR-bright stars and Sgr A* required an intermediate step, relying on the simultaneous radio detection of Sgr A* and masers associated with a subset of the IR-detected stars.
The angular offsets of IR-bright SiO maser-emitting stars from the $\sim$1~Jy Sgr A* continuum, along with the masers' proper
motions, provide an astrometric reference frame for the orbital solutions of stars much closer to the massive black hole than the maser-emitting stars \citep{menten1997,reid2003,reid2007,yelda2010,plewa2015,sakai2019,darling2023,hosek2025}.  This reference frame places
Sgr A* at rest and at the coordinate origin, which is also the phase center for the radio interferometry.

While the GRAVITY interferometer operating at ESO's VLTI can now detect Sgr A* in the near-IR continuum \citep{gravity2020},
it is still important to be able to refer IR positions to Sgr A* via the maser astrometry.
The GRAVITY science fiber FWHM is $\sim$74~mas in K-band \citep{gravity2021a}; stars at larger separations from Sgr A*
thus continue to rely on imaging data obtained at the adaptive optics resolution and referencing positions via the maser stars.
It is now also possible to address how well the positions of the mass (as measured by orbits), the near-IR emission of Sgr A*, and the radio emission of Sgr A* coincide.  The Event Horizon Telescope map of Sgr A* shows an asymmetry, and one assumes that the mass is located near the center (and not the centroid) of the luminous ring \citep{eht2022}.  Further, it is not obvious that the near-IR emission -- either in the quiescent state or during flares -- is centered on the mass.  It is also likely that the unresolved 43 GHz radio continuum does not coincide at some level with the IR emission or the mass; the
question is at what scale the standard assumptions break down.

As the reference frame continues to improve over time, both by way of better proper motion solutions from longer time baselines and from improved astrometric methods, the astrometry enables new science.  
The stars close to Sgr A* probe the black hole metric. So far, only effects of order $\beta^2$ ($\beta=v/c$), such as
Schwarzschild precession \citep{gravity2020}, have been detected.
These effects rely solely on the black hole mass, but in the next order in $\beta$, the equations of motion contain a term that is sensitive to black hole spin, commonly referred to as Lense-Thirring precession. Measuring this effect requires stars on orbits with very small pericenter distances.  The recent discovery of such a star, S-301, whose pericenter distance of 1.4 mas is roughly an order of magnitude smaller than that of S-02, could enable spin measurements of Sgr A* in the coming decades \citep{abd2026,piran2026,silva2026}.
It is therefore important to be able to pinpoint the black hole mass location for such close flybys.
At the fourth order, $\beta^4$, a quadrupole term enters the equations  of motion.  In general relativity, this term is fully determined by the black hole spin (assuming no charge), a consequence of the no-hair theorem \citep[e.g.,][]{will2008}. Hence, measuring both spin and quadrupole moment constitutes a test of fundamental physics, but it requires proper motion measurements below 10~$\mu$as~yr$^{-1}$ \citep{will2008}.
Applying analytical estimates from \citet{will2008} to the orbit of S-301, we estimate Lense-Thirring precession rates of roughly 6 $\mu$as~yr$^{-1}$ and quadrupolar precession rates of 0.2 $\mu$as~yr$^{-1}$ assuming a black hole spin parameter $\chi = 0.9$ \citep{daly2024}.

The stellar orbits are also sensitive to other masses in the region \citep{gravity2022}, and hence offer a possibility to probe the ``dark sector.''  The standard assumption is that this is dominated by stellar remnants, by far dominant over dark matter unless dark matter forms spikes. The presence of a cusp of stellar-mass black holes would inform science with the future gravitational wave experiment LISA
because gravitational waves from extreme-mass ratio inspirals would be produced from exactly such a cusp around a massive black hole \citep[e.g.,][]{hopman2006,alexander2009,amaro-seoane2011,babak2017}.
A cusp of pulsars would enable significant new tests of general relativity \citep{Liu2012,Hu2024}.
Conversely, a dark matter cusp would support the detection of axionic dark matter
\citep[e.g.,][]{hook2018,darling2020apj,darling2020prl,Foster2022,battye2022}.
 
\begin{figure}[t]
\begin{centering}
  \includegraphics[scale=0.45,trim= 30 30 20 10,clip=false]{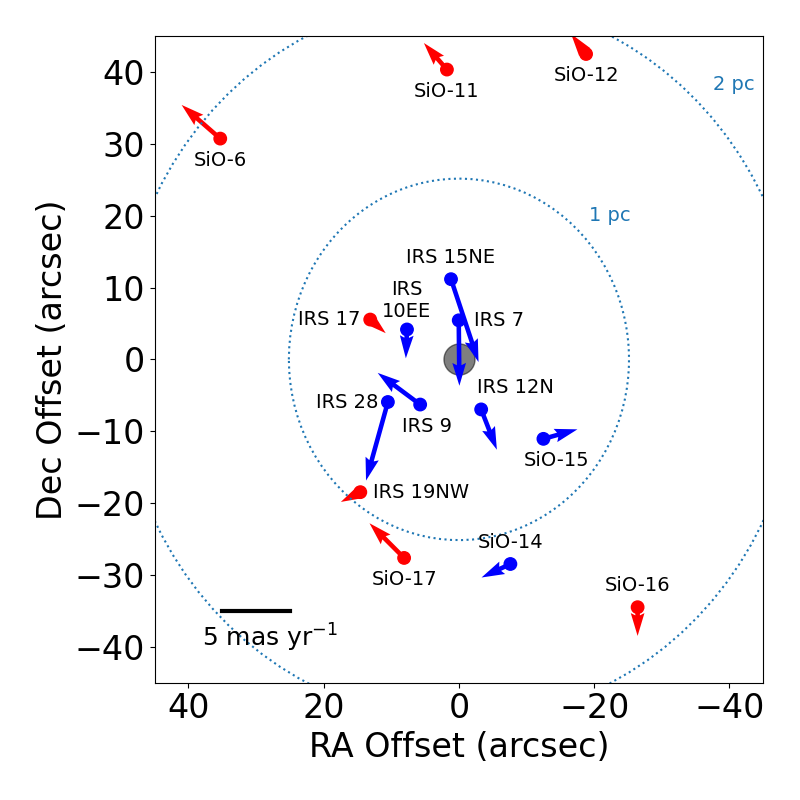}
  \caption{Epoch 2026.208 offsets and proper motions of the observed stellar SiO masers (Table \ref{tab:results2026} and Figure \ref{fig:pms}).
    Redshifted masers are red, and blueshifted masers are blue.  Sgr~A* is located at the grey circle (not to scale).
   Projected distances assume a Galactic Center distance of 8.2 kpc.}
\label{fig:map}
\end{centering}
\end{figure}

Astrometric SiO maser observations have been ongoing for nearly three decades.  While proper motions and the reference frame stability
improve over time as $t^{-3/2}$, assuming regular sampling, the flip side of this benefit is that astrometric solutions that are not updated
will decay as time passes because the uncertainty cone grows over time.
In this work, we update the reference frame and estimate its future decay.  
Section \ref{sec:data} describes the data sources, observations, and reduction used to form a 30-year astrometric solution for
15 SiO stellar masers.  Section \ref{sec:methods} outlines the astrometric methods that are described in detail in \citet{darling2023}.
Section \ref{sec:results} presents the improved astrometry and highlights the substantial variability in the maser spectra.  We discuss
the implications of the new reference frame and its decay over time in Section \ref{sec:discussion}, and we compare the maser-based
reference frame to the \citet{hosek2025} HST-Gaia reference frame in Section \ref{sec:analysis}.  Section \ref{sec:conclusions}
summarizes this work and discusses the future outlook for precise astrometry in the Galactic Center.

When physical transverse distances or transverse velocities are quoted, we assume the distance to Sgr~A* is 8.2~kpc
\citep[e.g.,][]{do2019,reid2019,gravity2021,leung2023}.


\section{\label{sec:data} Observations and Data Reduction}

This work adds a 2026 Karl G. Jansky Very Large Array (VLA)\footnote{The National Radio Astronomy Observatory
  is a facility of the National Science Foundation operated under cooperative agreement by Associated Universities, Inc.}
epoch to the 1996--2022 SiO stellar maser time series to form a 30-year baseline.
\citet{darling2023} describe the legacy data sources in detail, and their
properties are summarized in Table \ref{tab:data}.  As before, we focus on the subset of the stellar masers
and the maser transitions that have legacy astrometry ($J=1-0$, $v=1$ and $v=2$).
Table \ref{tab:results2026} lists the masers, and Figure \ref{fig:map} shows their locations and motions obtained in
Section \ref{sec:methods}.

\begin{deluxetable}{clccc}
  \tablecaption{\label{tab:data} SiO Maser Data and Observations}
  \tablehead{ \colhead{Mean Epoch} & \colhead{Program} & \colhead{$v$\tablenotemark{a}} & \colhead{Beam} & \colhead{Ref.} \\
    & & & (mas)}
  \startdata
1996.413  & VLBA  BM060  & 1 & $1.2\times0.9$  & R03 \\                
1998.410  & VLA AM592 & 1 & $70\times30$ & R03 \\          
2000.850  & VLA AR451 & 1 & $80\times40$ & R03 \\
2006.200  & VLA AR588 & 1 &  $86 \times 33$ & R07\\
2008.860  & VLA AR678 & 1 & $82\times35$ & R22 \\
2011.470  & VLA 11A-101 & 1 & $97\times42$ & R22 \\
2014.249 & VLA 14A-168 & 1,2 & $66\times30$ & R22 \\
2020.988 & VLA 19A-310 & 1,2 &$93\times36$ & P22,D23 \\
2022.227 & VLA 22A-328 & 1,2 & $82\times39$ & D23 \\
2026.208 & VLA 26A-049 & 1,2 & $81\times38$ & D26 \\
  \enddata
  \tablerefs{R03 = \citet{reid2003}; R07 = \citet{reid2007};
    R22 = Reid (2022, private communication); P22 = \citet{paine2022};
    D23 = \citet{darling2023}; D26 = this work.}
  \tablenotetext{a}{Vibration quantum number (all transitions are $J=1-0$).}
\end{deluxetable}

VLA program 26A-049 executed in three 5-hour sessions on 09, 20, and 23 March 2026.  This brings the time baseline to 29.795 years for a subset of the masers.
We observed the field of view centered on Sgr A* in the $v=1$ and $v=2$ 43 GHz SiO maser $J=1-0$ transitions.
The observations were identical to
previous program 22A-328 in order to minimize systematics and harmonize the time series.  Two SiO masers, the
43.1221~GHz $v=1$ $J=1-0$ and the 42.8206~GHz $v=2$ $J=1-0$ transitions, were observed simultaneously in 128~MHz
bandwidths with 100~kHz (0.7~km~s$^{-1}$) channels, two circular polarizations, and 3~s integrations using the largest VLA configuration.
Sgr A* served as the $\sim$1~Jy in-beam complex gain calibration source, 3C286 provided flux calibration, and J1924$-$2914 was
used for bandpass and delay calibration.  
Calibration, imaging, and astrometric measurements replicate those presented in \citet{darling2023}.  In short, we use
CASA \citep{CASA} for all steps.  We average two channels and two integrations prior to calibration, and the calibration
process is standard for Q-band A configuration.  We rely on per-integration complex gain calibration to track phase wrapping.

The Sgr A* continuum
provides the phase center to which all masers are referenced.  Sgr A* shows a significant proper motion with respect to
background quasars due to the barycenter motion about the Galactic Center
\citep{reid2020,xu2022,lewis-merrill2026}, so only relative astrometry is obtained from this method.  All maser
positions and subsequent proper motions are tied to the Sgr A* radio continuum, and it is assumed that this continuum
is fixed with respect to the massive black hole.  

In order to retain the relative in-beam astrometry, no continuum subtraction was performed, and we simultaneously imaged
Sgr~A* and all of the masers as outlier fields.  All three 26A-049 sessions were combined for imaging.  We used CASA \texttt{tclean}
to clean and image down to 15~mJy, five times the per-channel rms noise.   We measure Sgr A* and maser positions in the image plane
at every channel in each bandpass from 2D Gaussian fits using CASA \texttt{imfit}.   A variance-weighted average of peak fluxes $>5\sigma$
forms the final position and its uncertainty.  Offsets from the Sgr A* continuum centroid are calculated for each maser and added to the time series.  
Maser coordinate uncertainties are generally less than 1~mas (8 AU at 8.2 kpc).

\begin{figure*}[ht]
\begin{centering}
  \includegraphics[scale=0.48,trim= 20 36 20 20,clip=false]{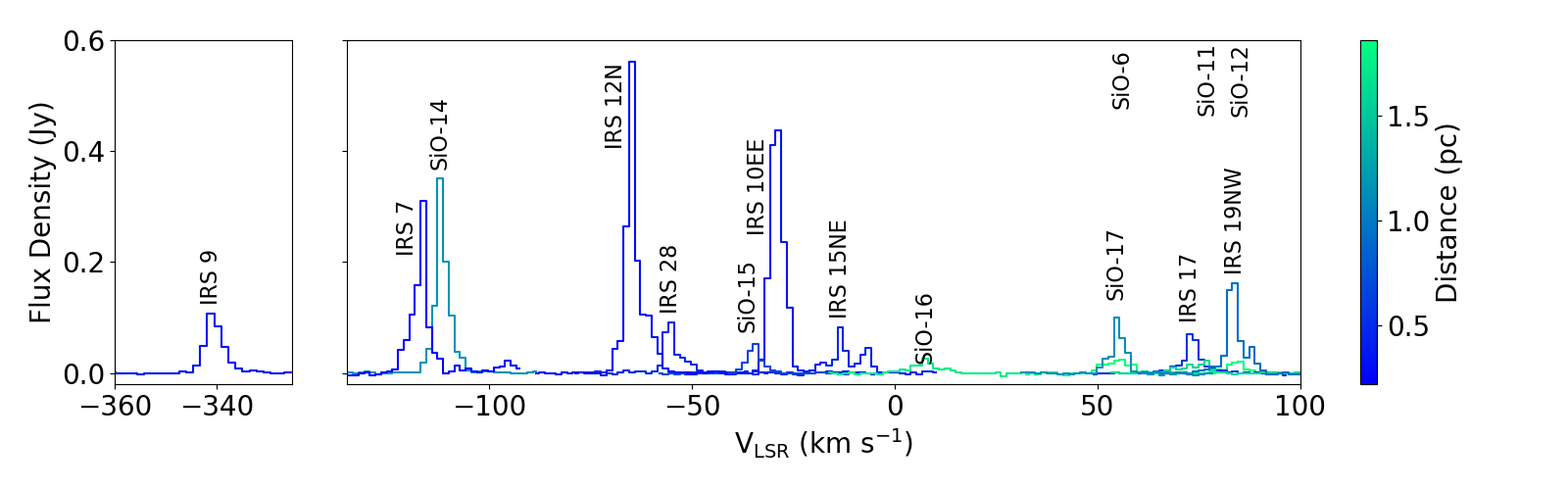}
  \includegraphics[scale=0.48,trim= 20 28 20 20,clip=false]{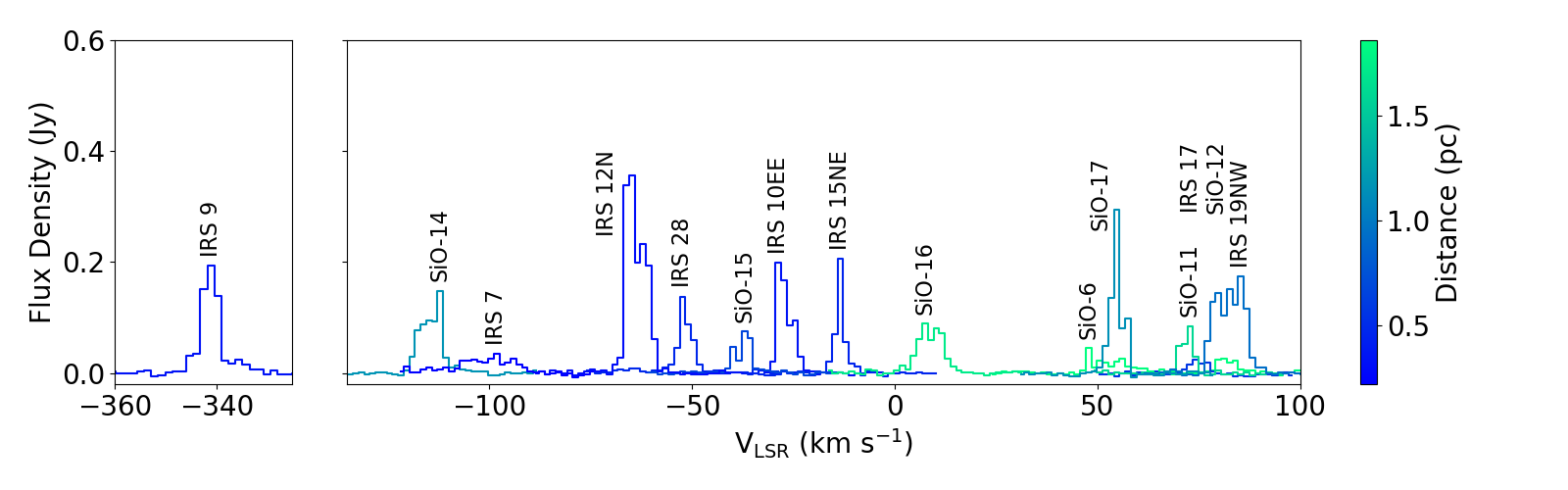}
  \caption{Composite spatially-integrated SiO $v=1$, $J=1-0$ 43.1221~GHz maser spectra from the
    2026.208 (top) and 2022.227 (bottom) image cubes.
    The colors indicate the projected distance from Sgr~A*, assuming a Galactic Center distance
    of 8.2 kpc.  The same flux density scale is used in both plots to show the variation between epochs.}
\label{fig:spectra}
\end{centering}
\end{figure*}

\begin{deluxetable*}{lrrrccrcc}
  \tablecaption{\label{tab:results2026} SiO Maser Angular Offsets and Proper Motions}
  \tablehead{ \colhead{Name} & \colhead{$v_{\rm LSR}$} & \colhead{RA Offset\tablenotemark{a}} & \colhead{Dec Offset} & \colhead{PM RA} &  \colhead{PM Dec} & \colhead{$\chi^2_\nu$} & \colhead{Ref.\ Epoch} & \colhead{$N_{\rm Obs}$}\\
 &  (km s$^{-1}$) & \colhead{ (arcsec)} & \colhead{(arcsec)} & (mas yr$^{-1}$) &  (mas yr$^{-1}$)  }
  \startdata
IRS 9  &  $-$341 &  $+5.71323 \pm 0.00007$  &  $-6.30393 \pm 0.00019$ &  $+3.056 \pm 0.013$ &  $+2.313 \pm 0.027$ & 0.8 & 2019.021 & 9 \\
IRS 7  &  $-$114 &  $+0.03302 \pm 0.00030$  &  $+5.47097 \pm 0.00058$ &  $-0.048 \pm 0.041$ &  $-4.584 \pm 0.073$ & 2.0 & 2017.836 & 7 \\
SiO-14  &  $-$111 &  $-7.61660 \pm 0.00016$  &  $-28.47017 \pm 0.00033$ &  $+2.151 \pm 0.025$ &  $-0.859 \pm 0.045$ & 3.3 & 2021.121 & 9 \\
IRS 12N  &  $-$65 &  $-3.27897 \pm 0.00008$  &  $-6.95064 \pm 0.00011$ &  $-1.099 \pm 0.014$ &  $-2.800 \pm 0.018$ & 1.9 & 2021.167 & 10 \\
IRS 28  &  $-$54 &  $+10.49314 \pm 0.00019$  &  $-5.87501 \pm 0.00031$ &  $+1.470 \pm 0.028$ &  $-5.440 \pm 0.048$ & 1.3 & 2015.418 & 9 \\
SiO-15  &  $-$35 &  $-12.47437 \pm 0.00021$  &  $-11.06500 \pm 0.00038$ &  $-2.486 \pm 0.035$ &  $+0.736 \pm 0.061$ & 1.4 & 2020.150 & 7 \\
IRS 10EE  &  $-$28 &  $+7.68493 \pm 0.00009$  &  $+4.16859 \pm 0.00012$ &  $+0.051 \pm 0.016$ &  $-2.027 \pm 0.016$ & 2.7 & 2021.531 & 10 \\
IRS 15NE  &  $-$11 &  $+1.20200 \pm 0.00014$  &  $+11.24489 \pm 0.00021$ &  $-1.902 \pm 0.014$ &  $-5.832 \pm 0.020$ & 1.0 & 2011.314 & 10 \\
SiO-16  &  $+$7 &  $-26.41836 \pm 0.00060$  &  $-34.47080 \pm 0.00110$ &  $+0.177 \pm 0.082$ &  $-1.793 \pm 0.149$ & 13.4 & 2017.293 & 8 \\
SiO-6  &  $+$52 &  $+35.26470 \pm 0.00078$  &  $+30.68813 \pm 0.00174$ &  $+2.581 \pm 0.079$ &  $+2.280 \pm 0.167$ & 8.0 & 2012.939 & 9 \\
SiO-17  &  $+$53 &  $+8.08969 \pm 0.00025$  &  $-27.65320 \pm 0.00043$ &  $+2.424 \pm 0.035$ &  $+2.652 \pm 0.057$ & 4.7 & 2017.666 & 8 \\
SiO-11  &  $+$71 &  $+1.76147 \pm 0.00053$  &  $+40.30656 \pm 0.00105$ &  $+1.661 \pm 0.080$ &  $+1.758 \pm 0.154$ & 23.3 & 2014.625 & 9 \\
IRS 17  &  $+$74 &  $+13.12283 \pm 0.00032$  &  $+5.54113 \pm 0.00064$ &  $-1.193 \pm 0.068$ &  $-0.990 \pm 0.112$ & 2.8 & 2024.842 & 7 \\
SiO-12  &  $+$82 &  $-18.80635 \pm 0.00063$  &  $+42.47991 \pm 0.00129$ &  $+1.196 \pm 0.117$ &  $+1.178 \pm 0.207$ & 4.1 & 2016.751 & 8 \\
IRS 19NW  &  $+$84 &  $+14.57844 \pm 0.00023$  &  $-18.47420 \pm 0.00046$ &  $+1.356 \pm 0.049$ &  $-0.665 \pm 0.084$ & 8.4 & 2020.946 & 9 \\
\enddata
\tablecomments{Coordinate offsets are with respect to the Sgr~A* radio centroid at the reference epoch, which is the position variance-weighted date of the time series (see Section \ref{sec:methods}).  The LSR velocity is
  approximate; the masers show variability in their spectral peaks and velocity centroids \citep{reid2003,reid2007,paine2022}.  The $\chi^2_\nu$ statistic characterizes the joint weighted least-squares proper motion fit in both coordinates. }
\tablenotetext{a}{This offset is corrected for Declination:  it is $\Delta {\rm RA} \cos({\rm Dec})$.}
\end{deluxetable*}

\section{\label{sec:methods} Astrometric Methods}

This work replicates the methods presented in \citet{darling2023}.  However, we do not apply time-dependent differential offsets to the
astrometry in each epoch, which leads to lower scatter in the time series.  
Table \ref{tab:time_series} lists the coordinate offsets from the Sgr A* continuum for all epochs.  

Proper motions $(\mu_\alpha,\mu_\delta)$ are obtained from linear fits to the offset-corrected coordinate time series
$(\Delta \alpha^*(t_{\rm obs}),\Delta \delta(t_{\rm obs}))$\footnote{$\Delta \alpha^* = \Delta \alpha \cos \delta$}
centered at a reference epoch $t_{\rm ref}$:
\begin{eqnarray}
  \Delta \alpha^* (t_{\rm obs})  & = &\Delta \alpha^*_{\rm ref} + \mu_\alpha (t_{\rm obs} - t_{\rm ref})\label{eqn:dra}, \\
  \Delta \delta(t_{\rm obs}) & = &\Delta \delta_{\rm ref} + \mu_\delta (t_{\rm obs} - t_{\rm ref}) \label{eqn:ddec}.
\end{eqnarray}
$t_{\rm ref}$ is the the variance-weighted mean date in the time series, where the variance is the sum in quadrature
of the coordinate uncertainties of each epoch.  This reference epoch selection method minimizes correlation between
the proper motion and the Sgr A* offsets at $t_{\rm ref}$, $(\Delta \alpha^*_{\rm ref},\Delta \delta_{\rm ref})$.
There is one reference epoch per maser, and both coordinate time series are fit simultaneously
to assess cross-coordinate correlations.  A quadratic term in Equations \ref{eqn:dra} and \ref{eqn:ddec}
could be added to assess transverse acceleration \citep{paine2022}.  

We re-examined the various time series fitting methods described in \citet{darling2023}.
As before, a variance-weighted least-squares linear fit in each coordinate is generally consistent with the other more
elaborate methods.  A few examples are depicted in Figures \ref{fig:PMresids} and \ref{fig:PMresidsEU}.
We employ joint RA and Dec weighted linear least-squares fits for all subsequent results and analysis.  
The reference epoch for each maser is the Sgr A* offset variance-weighted mean date in the time series, $t_{\rm ref}$, listed in Table
\ref{tab:results2026}; joint coordinate fits provide a single reference epoch rather than one for each coordinate solution.  
This can induce a minor correlation between the slope and intercept of single-coordinate fits, but the correlation between RA and Dec solutions is generally negligible.

We no longer assign an uncertainty floor to the IRS~7 offset from Sgr A*, which was done by \citet{reid2007} and
\citet{paine2022} owing to its maser variability and supergiant luminosity.  As described in Section \ref{sec:results} and shown in
Figures \ref{fig:error_vs_time} and \ref{fig:pms} and Table \ref{tab:results2026}, its astrometric time series shows less scatter
than many other masers. We do, however, continue to restrict the time series to 2006.200 and later \citep{darling2023}.

\citet{li2010} and \citet{paine2022} have identified additional masers in the inner few parsecs around Sgr A* that could in principle contribute
to the Galactic Center reference frame.  
While observations in 2019, 2022, and 2026 provide an adequate time series to incorporate new masers into the reference frame solution,
these masers are more distant in projection from Sgr A* than the primary sample and are typically of low flux density.  As a result, they would not significantly improve the signal-to-noise-weighted astrometry.  They are therefore omitted.

\section{\label{sec:results} Results}

All masers were detected in the 2026 epoch, but substantial spectral variation is apparent.
Some masers were significantly brighter than in previous sessions, some were fainter, and many changed
their shapes and velocity centroids.
Figure \ref{fig:spectra} compares the 2026 epoch with the 2022 observations of the $v=1$, $J=1-0$ 43.1221~GHz transition.
Most notable is the 30-fold flare in the very blue component of IRS~7:  this component was not detected in 2022 but was
luminous in pre-2000 observations \citep{reid2003,reid2007}.
While the maser light centroid, which is dominated by the blue component in 2026,
  is not anomalous compared to recent epochs  (see Section \ref{subsec:2022} and Figure \ref{fig:pms}), 
the maser velocity and line shape changes will ultimately limit line of sight acceleration
measurements compared to transverse accelerations \citep{paine2022}.  

Table \ref{tab:results2026} lists the new Sgr A* offsets, proper motions, reference epochs, and number of epochs used in the proper motion fits.  
Figure \ref{fig:pms} shows the maser proper motion fits and residuals, and Table \ref{tab:time_series} shows all of the data that are depicted in Figure \ref{fig:pms}.

The aggregate statistics of the astrometric solutions have improved compared to the \citet{darling2023} results (see Section
  \ref{sec:analysis}).  
The uncertainties in the variance-weighted means of the maser coordinate offsets from Sgr A* are $(0.06, 0.09)$~mas in the current epoch.  These correspond to
$(0.5,0.7)$~AU, which may be smaller than the SiO-emitting region in the host stars.
The maser emission is often asymmetric around the stellar center of mass and variable in time, so the size of the unresolved maser-emitting region may be a floor on the astrometric accuracy for tracking the systemic stellar position.
The mean maser proper motion uncertainties are $(0.05,0.08)$~mas yr$^{-1}$ in RA and Dec, corresponding to 
1.8 km s$^{-1}$ and 3.2 km s$^{-1}$.  The reference frame stability, the uncertainty in the variance-weighted mean proper motion difference between the IR\footnote{The IR proper motions are from \citet{sakai2019} measurements of seven maser-emitting stars (see Section
  \ref{subsec:sakai2019}).} and maser observations,
is $(10.2,14.7)$~$\mu$as~yr$^{-1}$  or $(0.40,0.57)$~km s$^{-1}$.
These ensemble characterizations of the reference frame have improved on the \citet{darling2023}
  results with the addition of the 2026 epoch (Section \ref{subsec:2022}).

\section{\label{sec:discussion}Discussion}

\begin{figure}[t]
\begin{centering}
  \includegraphics[scale=0.38,trim= 20 20 20 0,clip=false]{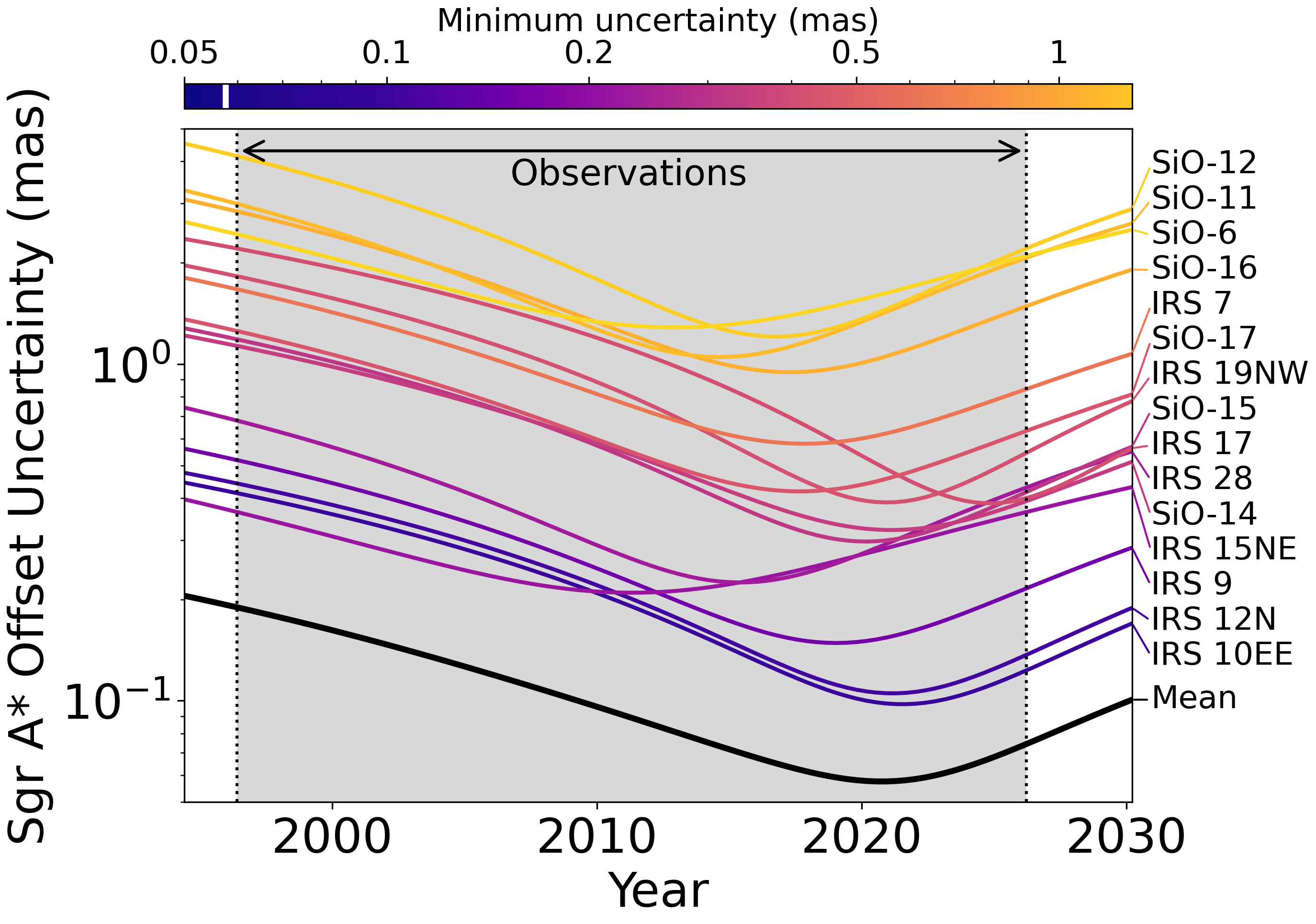}
  \caption{Maser position uncertainty vs.\ time.  This is the uncertainty in the offset from Sgr A* as a function of time based on the 30-year solution (Table \ref{tab:results2026}),
    combining Equations \ref{eqn:dra} and \ref{eqn:ddec}.  The black locus
    shows the uncertainty in the variance-weighted mean offset of the maser ensemble, and its minimum
     value is indicated by the white line in the colorbar.  
    The grey region indicates
    the 30-year span of observations of a subset of the masers.
    The increasing uncertainties beyond the observation window show how the astrometry decays into the future (and the past).  }
\label{fig:error_vs_time}
\end{centering}
\end{figure}

Given the astrometric solutions in Table \ref{tab:results2026}, one can compute the coordinate offsets of each maser from Sgr A* as a function of time as
well as the uncertainty in those offsets.  The uncertainties provide an estimate of the quality of the offsets versus observing epoch and a prediction for the
growth of the uncertainty cones as the astrometric solution decays.  
Figure \ref{fig:error_vs_time} shows the uncertainties in the radial offsets of the masers from Sgr A*.
The best-measured masers, IRS 10EE and IRS 12N, reach a minimum uncertainty of $\sim$0.1~mas.
The uncertainty  in the error-weighted mean offset of the maser ensemble shows a minimum
value of  0.06~mas in year 2020.9 and rises to 0.08~mas in the current epoch.

\begin{figure}[t]
\begin{centering}
  \includegraphics[scale=0.36,trim= 30 30 30 0,clip=false]{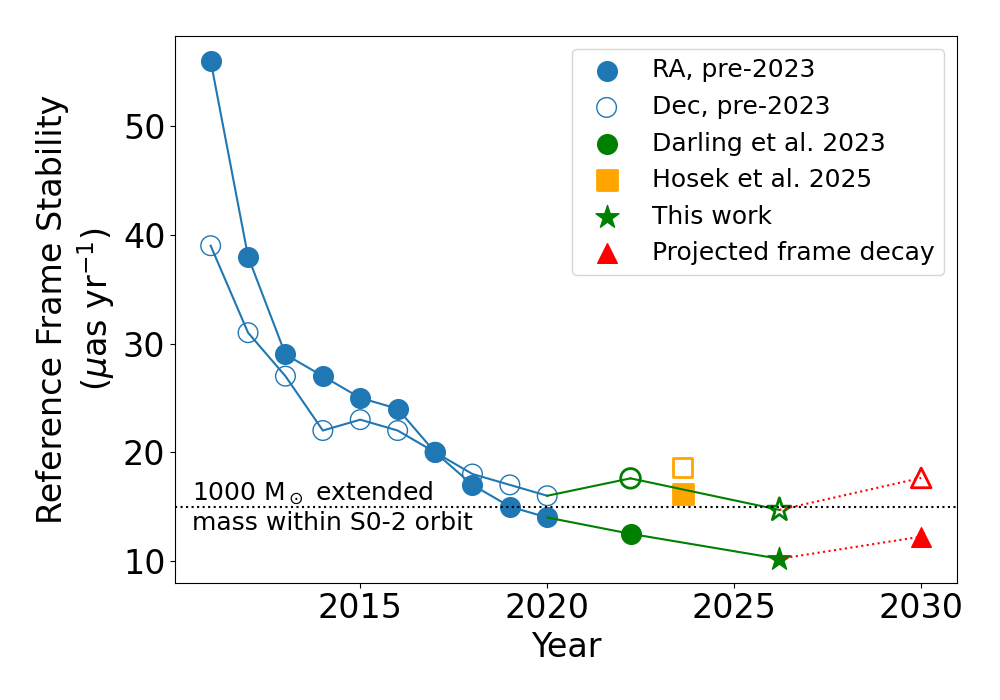}
  \caption{Reference frame stability vs.\ time.  The reference frame stability is
    the uncertainty in the variance-weighted mean of the proper motion differences between the IR and radio
    measurements of the stellar masers.  The blue and green symbols are referenced to the 
    IR astrometry of seven maser-emitting stars reported in \citet{sakai2019}.
  Filled points show the trend in RA, and the open symbols show Dec.  The red triangles indicate the projected
    reference frame stability decay in the absence of continued astrometric monitoring.
    The horizontal dotted line indicates the stability threshold needed to detect 1000~M$_\odot$ of mass external to Sgr A* but within
    the orbit of S0-2/S2 \citep{hosek2025}.}
\label{fig:stability}
\end{centering}
\end{figure}

The IR-maser reference frame stability has steadily declined (improved) over time.\footnote{The reference frame stability metric
used in \citet{darling2023} differs from the one used in this work.  When comparisons are made, such as in Figure \ref{fig:stability}, all previous values have been recalculated using a common (the current) definition.}
Figure \ref{fig:stability} shows this trend in RA since 2009 and in Dec since 2022.
The stability fell below a fiducial 15~$\mu$as~yr$^{-1}$ in RA in 2022 \citep{darling2023}
and in Dec in 2026 (this work).   This threshold corresponds roughly to the proper motion uncertainty needed to
use the S0-2/S2 stellar orbit to probe down to a mass of 1000~$M_\odot$ lying between its apocenter (0.01 pc)
and Sgr A*, assuming a 13~yr observation span \citep{hosek2025}.  Figure \ref{fig:stability} also shows the
reference frame stability between the masers and the \citet{hosek2025} HST-Gaia astrometry discussed in Section \ref{sec:analysis}.

The results of the latest epoch of observations, combined with those spanning the last 30 years, show that:
(1) the last 3--4 epochs have very stable astrometry compared to the full history (e.g., IRS 12N, IRS 10EE, SiO-12, and IRS 19NW in Figure \ref{fig:pms});
(2) there are, however, masers with persistent significant astrometric variations such as SiO-16, SiO-6, SiO-11, and IRS 19NW, indicated by large $\chi^2_\nu$ values
in Table \ref{tab:results2026}; 
(3) the proper motion uncertainties and reference frame stability continue to improve over time (no precision floor has yet been reached); and
(4) the reference frame should be durable into the early 2030s (Figure \ref{fig:stability}).
To maintain the stability and precision of the reference frame, the SiO maser observations should be repeated in roughly five years.

There are six masers that show significant variance far in excess of the measurement uncertainties ($\chi_\nu^2 > 4$).
Interestingly, these all show positive radial velocities, and the only positive radial velocity maser with low variance is IRS~17, which is also closest in projection to Sgr A* among this subset.
SiO-6 shows significant departures from the proper motion trendline in both coordinates with excursions as large as 10~mas.  It is not obviously periodic, and the angular
scale of the offsets is likely too large to be due to scintillation.  It may be noteworthy that the significant spectral change shown in Figure \ref{fig:spectra} is
accompanied by a significant astrometric change (Figure \ref{fig:pms} and Table \ref{tab:time_series}).  The per-channel uncertainty in each coordinate spans 1--8~mas
in recent epochs, and variation in the number and weighting of spectral channels could therefore alter the measured position.  Underestimated uncertainties
in the time series of SiO-6 are explored in Appendix \ref{sec:pms}.  
The large variance in SiO-16, SiO-17, SiO-12, and IRS 19NW is dominated by early epochs; all show astrometric convergence during the last 3--4 epochs, since 2014,
where the scatter is consistent with the measurement error or within the expected SiO-emitting region of the star.  
The SiO-11 astrometry is highly suggestive of periodicity, but a rough estimate suggests that
the binary companion would be at least 20~$M_\odot$.  
\citet{oyama2008} identified a similar trend over smaller timescales and smaller astrometric excursions for IRS 10EE, claiming a 13--16~$M_\odot$ companion
based on VLBA observations, but the VLA data lack the resolution to test this claim.

\section{\label{sec:analysis}Comparison to Previous Reference Frames}

We compare the current maser astrometry to previous work using the subset of masers that overlap each sample.  In each case, we quantify
the proper motion differences and the Sgr A* offset differences for a specific common reference epoch.
The 2026 maser epoch positions and proper motions are subtracted from those of 
each comparison reference frame -- not ordered by epoch -- in order to compare to the \citet{hosek2025} results as
well as for consistency across samples.

\subsection{IR AO Astrometry}\label{subsec:sakai2019}

\citet{sakai2019} present IR AO astrometry for seven maser-emitting IR-bright stars:
IRS 9, IRS 7, IRS 12N, IRS 28, IRS 10EE, IRS 15NE, IRS 17.  These are the stars used in Figure \ref{fig:stability}.
This IR astrometry relies on earlier maser astrometric epochs for the offsets from Sgr A* \citep{reid2007}.  
  The optimal common epoch is 2017.3339.  The error-weighted mean proper motion differences are
  $(+0.004 \pm 0.010, +0.041 \pm 0.015)$ mas~yr$^{-1}$ in RA and Dec; there is no significant ($>3\sigma$) difference between
  the two reference frames.  The weighted mean differences in offsets from Sgr A* are
  $(+0.043 \pm 0.079, +0.450 \pm 0.118)$ mas.  The Declination difference is significant, at $3.8\sigma$.  This is likely 
  attributable to the small sample whereby a single maser can significantly alter the mean.  Indeed, a bootstrap resampling
  of the stars produces much larger uncertainties in the mean values.  For the position offset differences, the uncertainties are
  estimated to be $(\pm0.25,\pm0.30)$~mas and the differences are not significant.
  Nonetheless, it is clear that a new IR-radio registration is needed.

\subsection{2022 Maser Astrometry}\label{subsec:2022}

  Comparison to the \citet{darling2023} astrometry includes all 15 masers, and the 2026 measurements are
  effectively a refinement of previous work.  Nevertheless, it is useful to assess the changes and to identify any masers
  that have significantly different positions or proper motions.  The weighted mean proper motion differences are 
  $(+1.8 \pm 10.3, +0.3 \pm 14.6)$ $\mu$as~yr$^{-1}$, i.e., the proper motions are consistent, and the differences are
  extremely small.
  The weighted mean Sgr A* offset differences in the common epoch 2014.7250 are $(-0.149 \pm 0.081, 0.126 \pm 0.129)$ mas
  and are not significant.  
  Individual masers do not show significantly different proper motions or Sgr A* offsets in the common epoch, except for
  IRS 9.  IRS 9 shows a 2.98$\sigma$ difference in RA offset, but it has the lowest postion uncertainty in the sample, equivalent
  to 1.1~AU.  We may be measuring changes in the maser light centroid at these scales, which is supported by the roughly
  factor of two change in brightness observed between 2022 and 2026 (Figure \ref{fig:spectra}).
  This phenomenon may set a future astrometric precision floor on additional masers.

\subsection{HST-Gaia Astrometry}

\begin{figure}[t]
\begin{centering}
  \includegraphics[scale=0.29,trim= 15 30 10 0,clip=false]{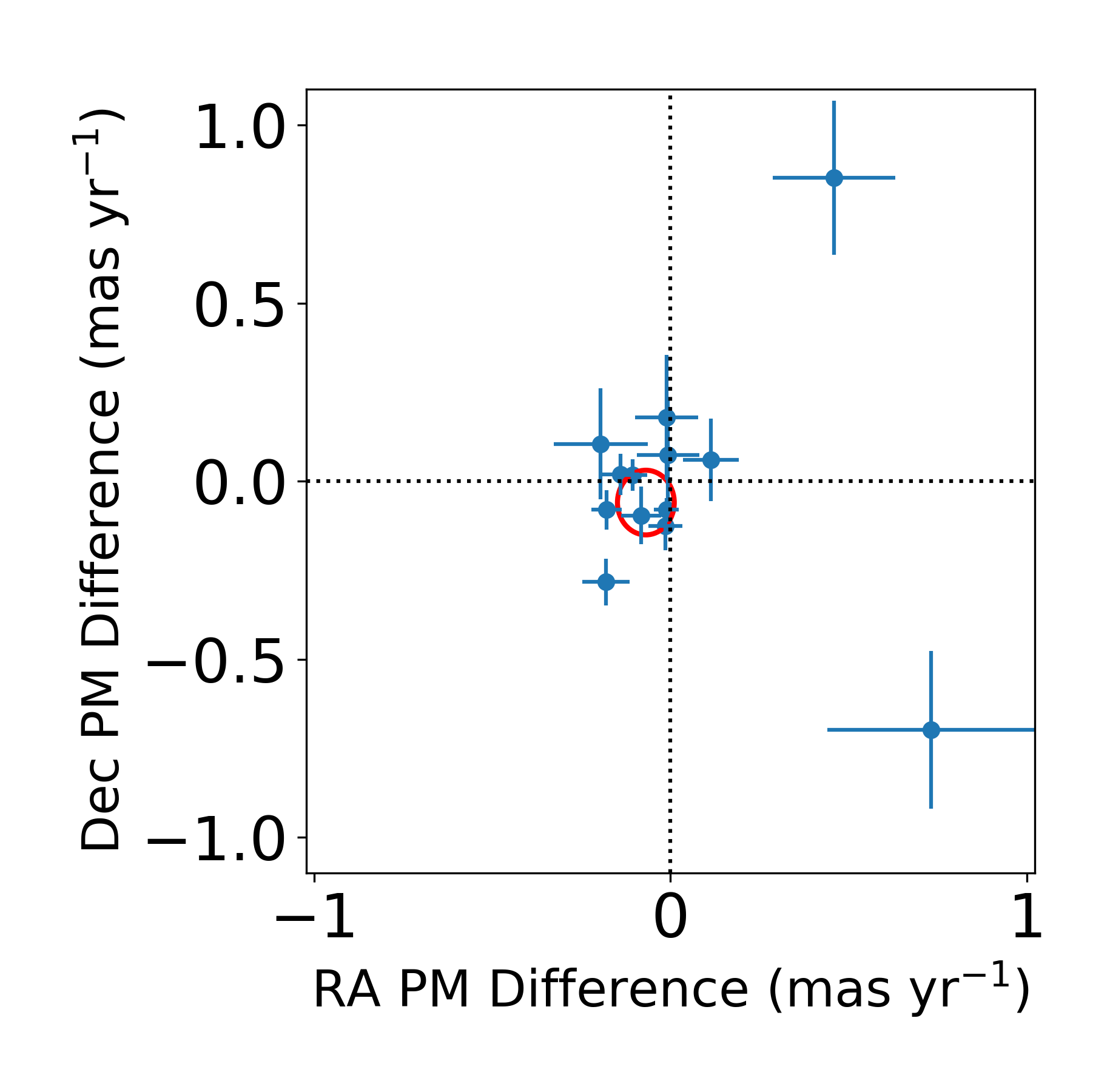}
  \includegraphics[scale=0.29,trim= 15 35 10 0,clip=false]{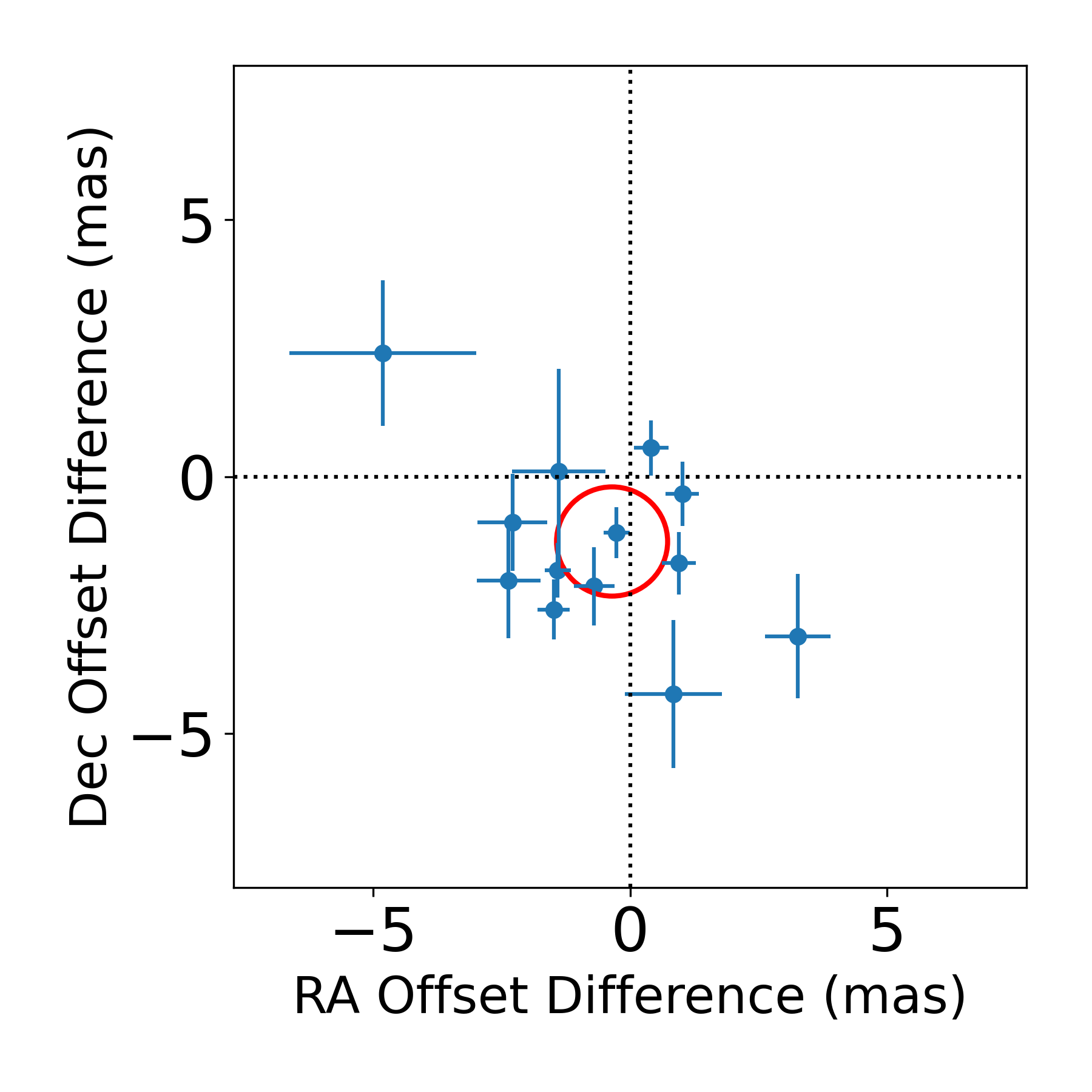}
  \caption{Difference between the \citet{hosek2025} HST-Gaia proper motions (left) and Sgr A* offsets (right)
      and the proper motions and offsets measured in this work.  The red ellipse indicates the 3$\sigma$ contour around the error-weighted mean based on bootstrap resampling of the 13 stars in the comparison.}
\label{fig:comparison}
\end{centering}
\end{figure}

Following the comparison of the \citet{hosek2025} HST-Gaia sample to the \citet{darling2023} stellar maser reference frame for 13 of the masers
in this work, we re-examine the differences in proper motions and positions of the overlap sample.\footnote{This treatment omits SiO-12 and SiO-17,}  Unlike the \citet{sakai2019} astrometry, the HST-Gaia astrometry is not referenced to any maser measurements and
represents an independent reference frame tied to the International Celestial Reference System (ICRS; \citealt{arias1995}).
We use the \citet{hosek2025} ``SgrA* at-rest'' coordinates and proper motions to match the maser measurements.
Based on Equations 10--12 of \citet{hosek2025}, the optimal common epoch is 2017.8145, the error-weighted difference in proper motions is
$(-0.068 \pm 0.013, -0.061 \pm 0.017)$ mas~yr$^{-1}$ in RA and Dec, and the error-weighted difference in coordinate offsets from Sgr A* is
$(-0.50 \pm 0.07, -1.15 \pm 0.10)$ mas.  This preserves the sign convention (subtracting the HST-Gaia values from the maser values) but
omits the systemic reference frame uncertainties between HST-Gaia and the maser reference frame.
If these uncertainties are included in quadrature, the differences become $(-0.069 \pm 0.016, -0.060 \pm 0.019)$ mas~yr$^{-1}$ and
$(-0.35 \pm 0.11, -1.26 \pm 0.20)$ mas.  These are consistent with the \citet{hosek2025} results, but with $\sim$50\% smaller uncertainties.

The differences in proper motion and offset between the stellar maser reference frame and the HST-Gaia frame
are all significant ($> 3\sigma$), especially the proper motion in RA (4.2$\sigma$) and Declination offset (6.3$\sigma$).
\citet{hosek2025} found $\sim$2.5$\sigma$ tension in the RA proper motion and position; these are now confirmed.
However, bootstrap resampling of the stars produces larger uncertainty estimates:
$(\pm0.027,\pm0.031)$ mas~yr$^{-1}$ in proper motion differences and  
$(\pm0.36,\pm0.35)$ mas in offset differences.  The bootstrap bias is negligible.
Based on the bootstrap, only the Dec offset remains significant, at 3.6$\sigma$.  
Figure \ref{fig:comparison} plots the proper motion and offset differences.

We note that the difference in RA coordinate offset shows a significant ($>3\sigma$) gradient with RA
across the field of view:  $-0.12 \pm 0.02$~mas per arcsec.  No other differential quantity shows a significant gradient in either coordinate direction.  
Given the small mean proper motion and offset statistical uncertainties in the maser sample described in Section \ref{sec:discussion} and depicted in
Figures \ref{fig:error_vs_time} and \ref{fig:stability}, it is important to identify the systematic source of the differences between reference frames.  The small number of maser-emitting stars is likely a factor, as demonstrated by the bootstrap uncertainty estimates.
Another factor is likely to be the systematic biases in the Gaia astrometric solution \citep[e.g.,][]{lindegren2021},
potentially exacerbated by the dearth of observable quasars near the Galactic Center.

\section{\label{sec:conclusions}Conclusions}

This work presents a 30-year baseline for the stellar SiO maser-based reference frame tied to the continuum of Sgr A*.  The weighted mean uncertainty in the maser
offsets from Sgr A* is 0.25 mas, which is similar to the size of the SiO-emitting regions (2.0 AU).  The mean proper motion uncertainties are $(0.05,0.08)$~mas~yr$^{-1}$
or $(1.8,3.2)$~km~s$^{-1}$ in RA and Dec.  The reference frame stability reaches sub-km~s$^{-1}$ levels: $(0.40,0.57)$~km~s$^{-1}$ or $(10.2,14.7)$~$\mu$as~yr$^{-1}$.
These represent a new level of precision, and the reference frame decay should be minor for $\sim$5 years but should be
refreshed in the early 2030s.  The statistical uncertainties in the reference frame, however,
show that systematic differences between this frame and the \citet{hosek2025} HST-Gaia frame must be reconciled.  

At some point, acceleration will be detected in the masers via proper motion curvature and/or changing radial velocities.  No credible accelerations have been detected to date, but the best limits are below $\sim$1~km~s$^{-1}$~yr$^{-1}$ \citep{paine2022}.  Neglecting accelerations will lead to growing errors in the reference frame, but including them will enable new and more precise physical measurements.  Future IR stellar tracking will use an accelerating reference frame, and it will be essential to match this to the masers.
Acceleration measurement also determines a stellar maser's physical distance from Sgr A* as well as the
  star's orbital parameters.  If
acceleration is not detected, the acceleration upper limit corresponds to a distance lower limit.  The inherent motion of the
maser-emitting shell (which is detectable in variable maser spectra and is a source of noise in astrometry) may be confused with systemic acceleration of the star over short time scales, but any acceleration detected over several decades should reflect the stellar kinematics.

\acknowledgments
We dedicate this work to Karl Menten, a pioneer in masers, molecules, astrometry, and radio astronomy in general.
We thank Mark Reid for sharing his expertise and guidance and for passing this interesting baton to the next generation,
and we thank the anonymous referee for helpful feedback.
JD acknowledges support from NSF grant AST-1908122. M.W.H. is supported by the Brinson Prize Fellowship.
  This research made use of NumPy \citep{NumPy}, Matplotlib \citep{Matplotlib}, and
  Astropy\footnote{\url{http://www.astropy.org}}, a community-developed core Python package for Astronomy \citep{astropy:2013, astropy:2018}.

  \facility{VLA, VLBA}
  \software{CASA \citep{CASA}, astropy \citep{astropy:2013, astropy:2018}, NumPy \citep{NumPy}, Matplotlib \citep{Matplotlib},
    \texttt{lmfit} \citep{newville2021}, \texttt{emcee} \citep{emcee}, CARTA \citep{carta2021} }

\clearpage

\bibliography{ms}

\appendix

\restartappendixnumbering

\section{Astrometric Offset Time Series}\label{sec:time_series}

Table \ref{tab:time_series} lists the Sgr A* coordinate offset time series for each maser used in this analysis.  Figure \ref{fig:pms} shows the
time series in each coordinate, the least-squares proper motion fit residuals, and the time series on the sky.

\begin{figure*}[t]
\begin{centering}
  \includegraphics[scale=0.30,trim= 20 20 20 0,clip=false]{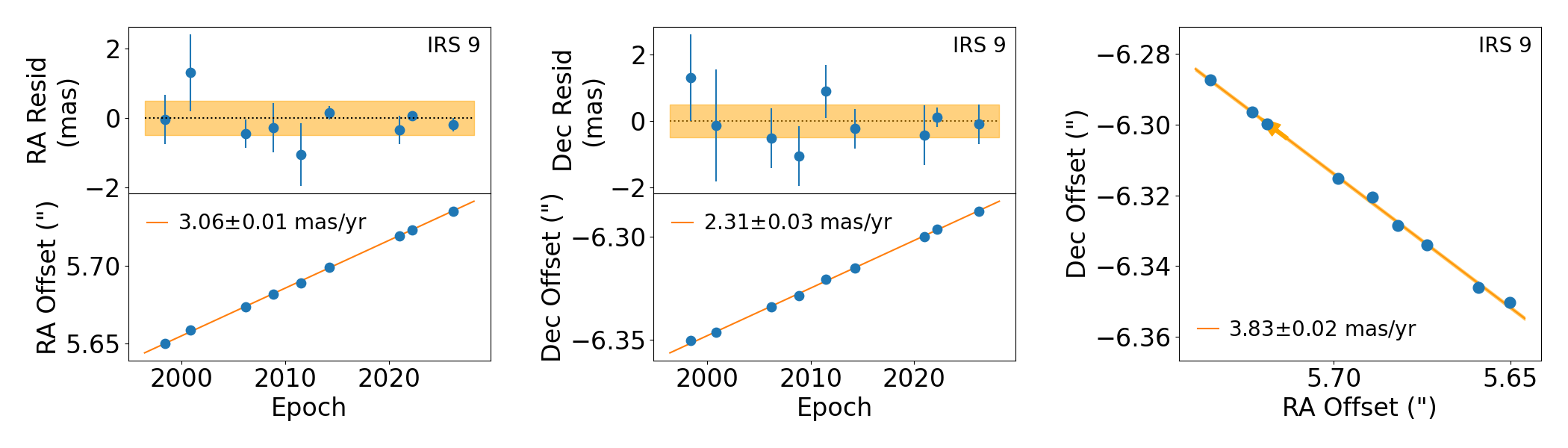}
  \includegraphics[scale=0.30,trim= 20 20 20 0,clip=false]{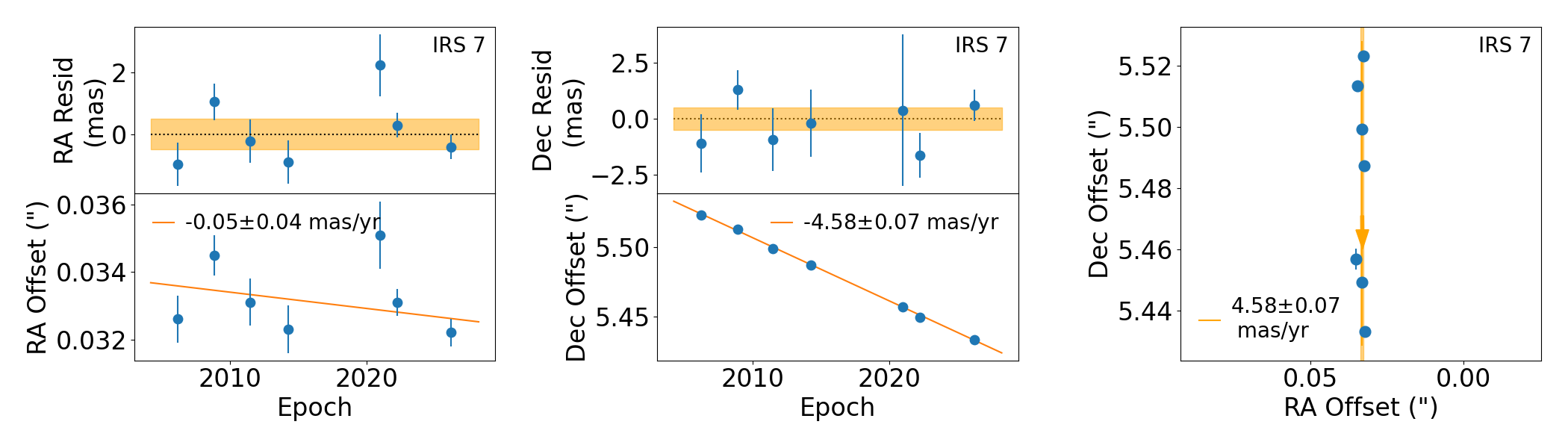}
  \includegraphics[scale=0.30,trim= 20 20 20 0,clip=false]{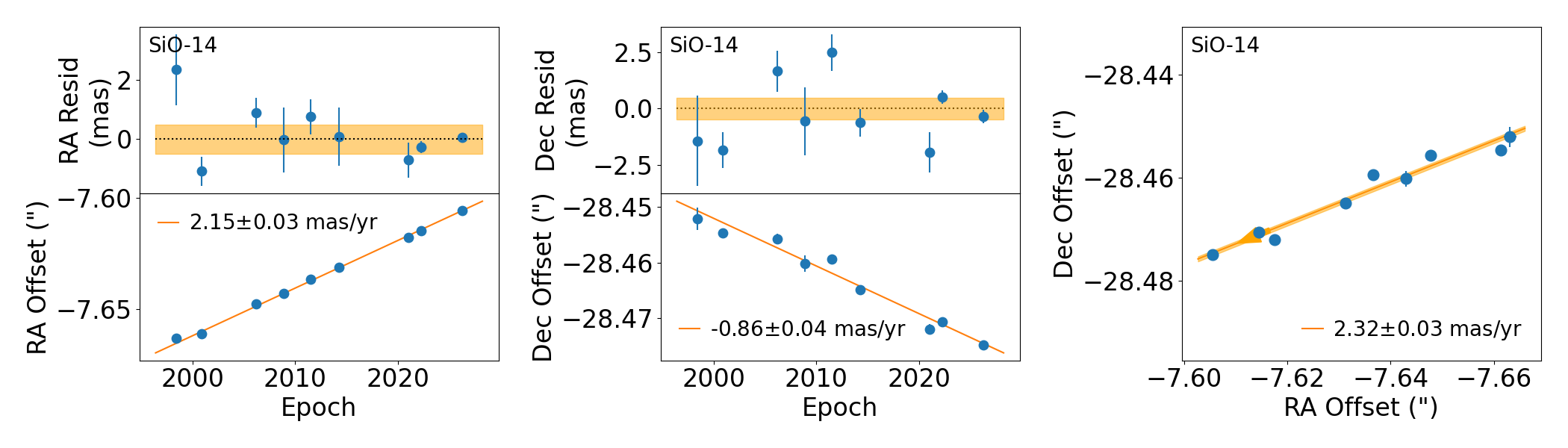}
  \includegraphics[scale=0.30,trim= 20 20 20 0,clip=false]{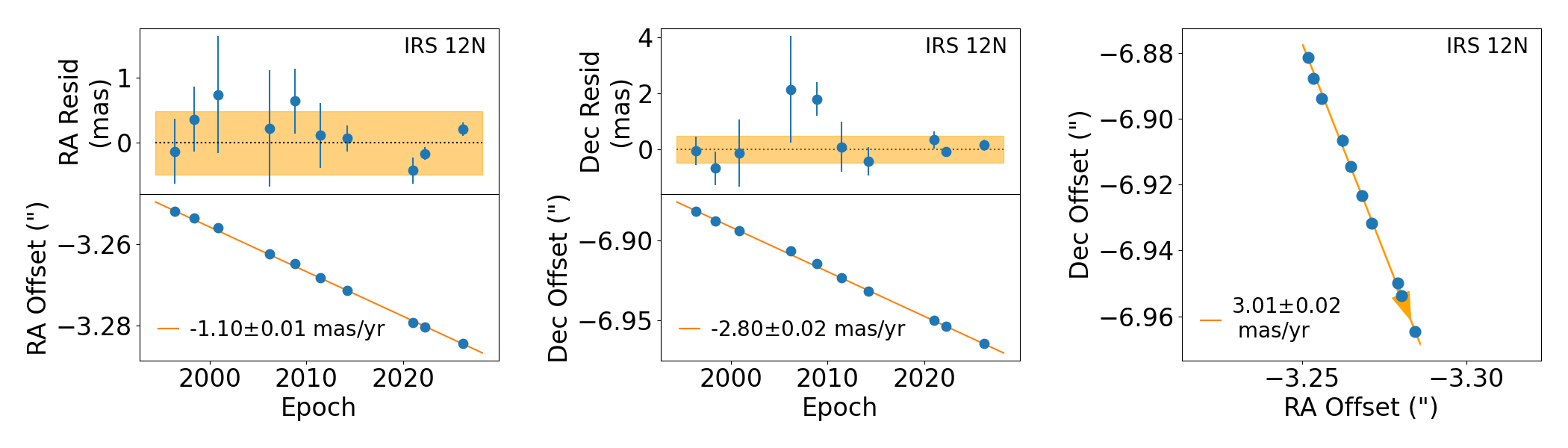}
  \includegraphics[scale=0.30,trim= 20 20 20 0,clip=false]{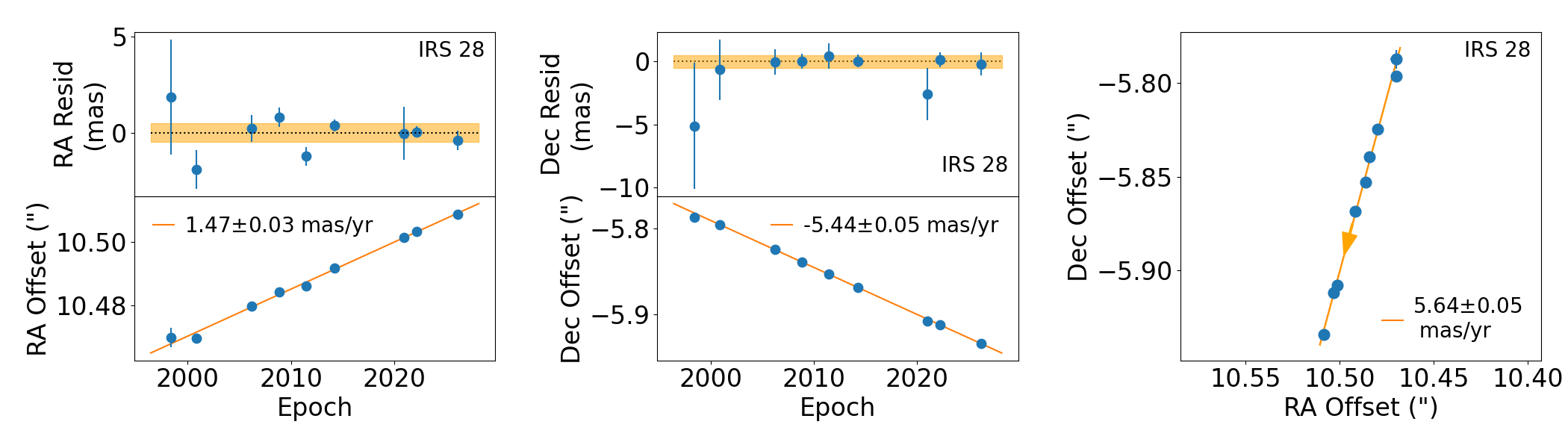}
  \caption{Time series for each stellar maser, as an offset from Sgr A* (rows).  We show the motion on the sky (right column; the arrows indicate the
    direction of proper motion), the time series in RA (left), and the time series in Dec (center).  For each coordinate time series, the orange
    lines show the least-squares proper motion fits, and the upper panels show the time series residuals.  The fit line width indicates $\pm4$~AU in all panels.  The time series is connected sequentially when it is ambiguous, as in SiO-16.  RA offsets are corrected for $\cos({\rm Dec})$.}
\label{fig:pms}
\end{centering}
\end{figure*}

\addtocounter{figure}{-1}

\begin{figure*}[t]
\begin{centering}
  \includegraphics[scale=0.30,trim= 20 20 20 0,clip=false]{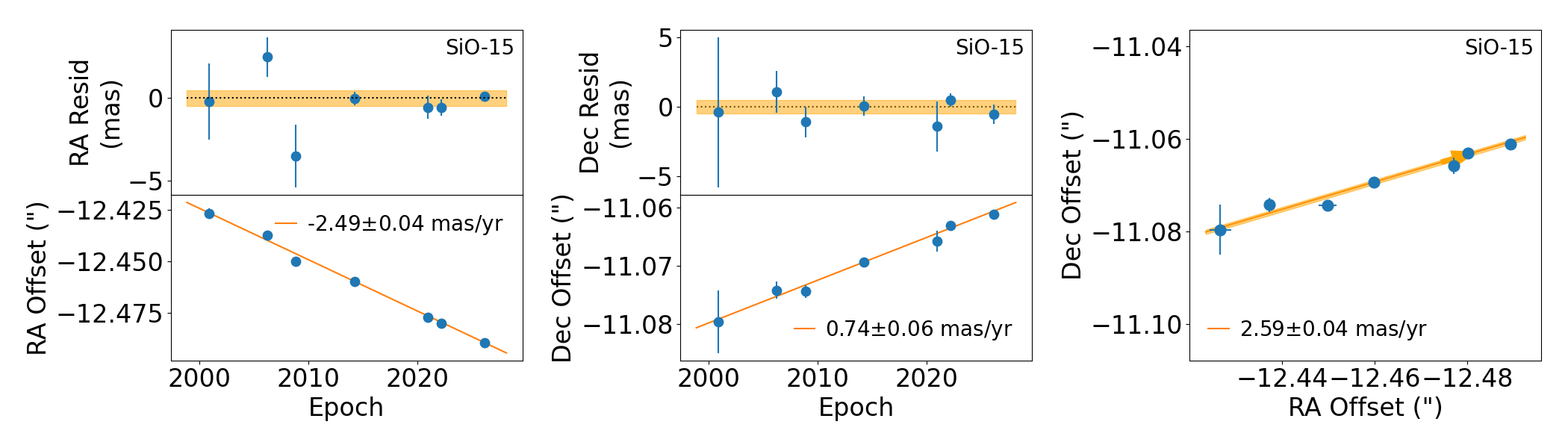}
  \includegraphics[scale=0.30,trim= 20 20 20 0,clip=false]{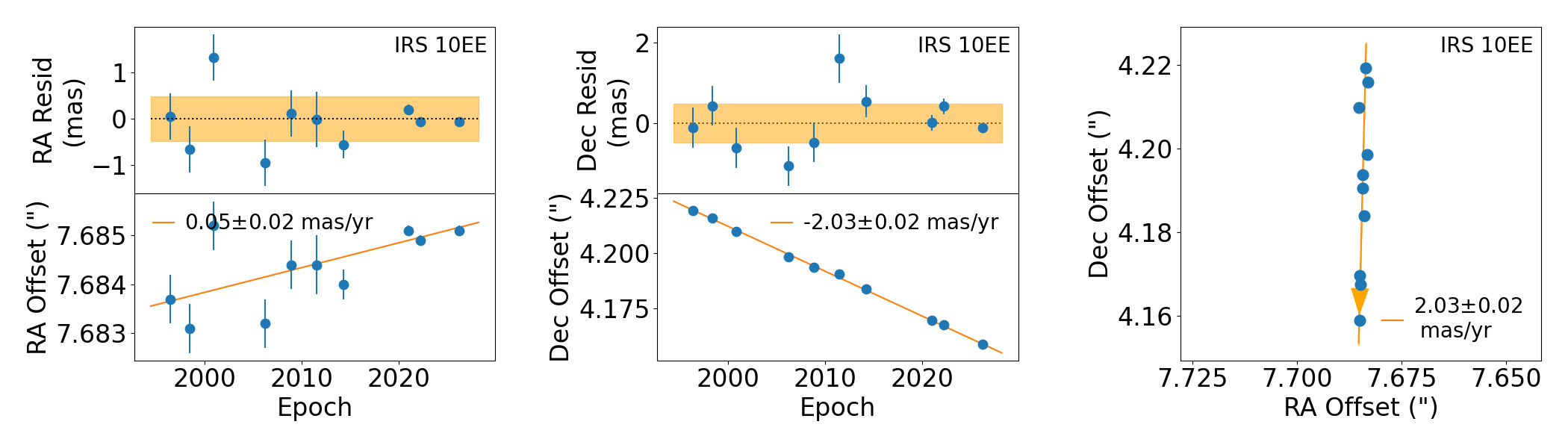}
  \includegraphics[scale=0.30,trim= 20 20 20 0,clip=false]{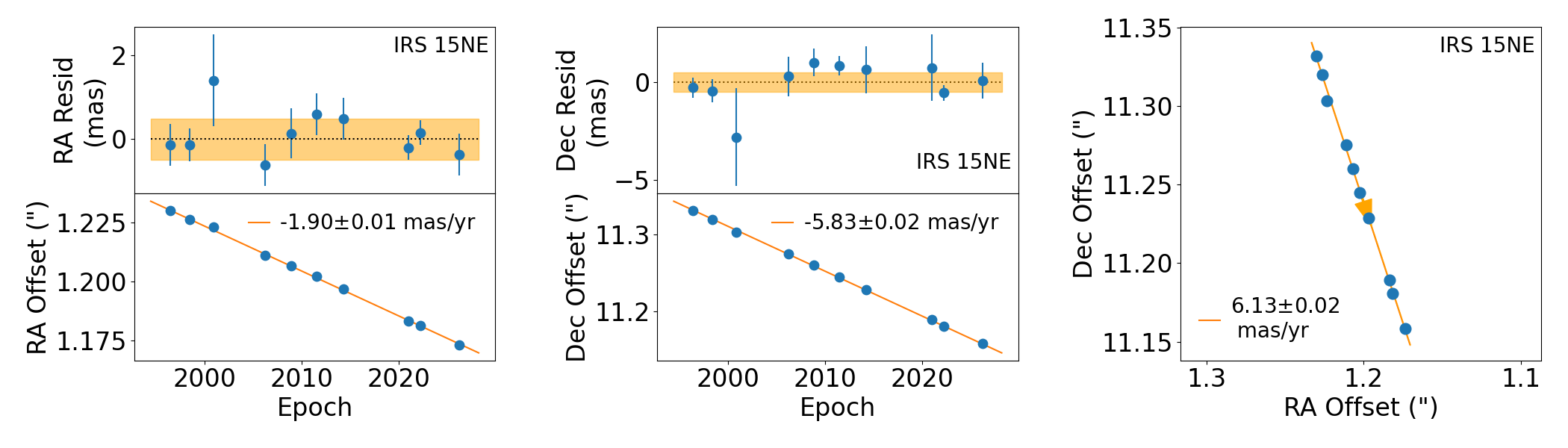}
  \includegraphics[scale=0.30,trim= 20 20 20 0,clip=false]{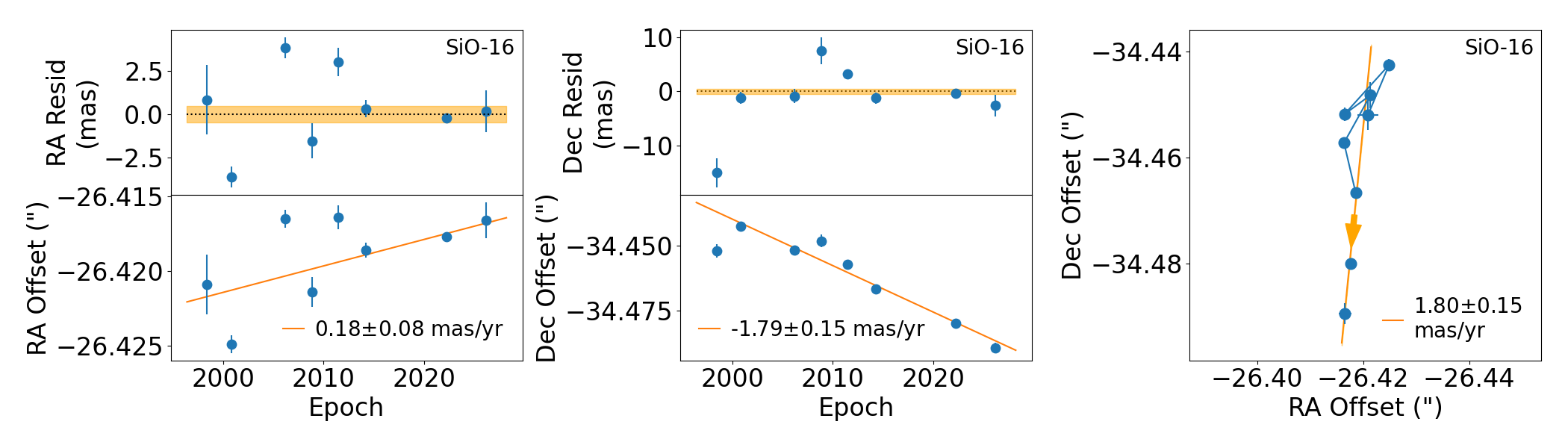}
  \includegraphics[scale=0.30,trim= 20 20 20 0,clip=false]{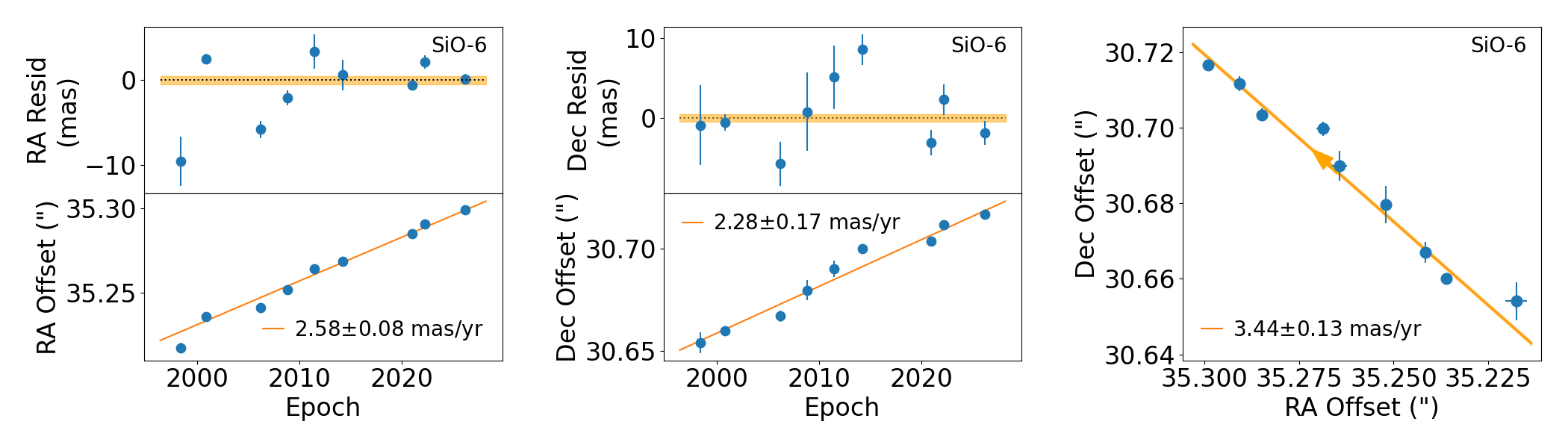}
  \caption{{\it continued}}
\end{centering}
\end{figure*}

\addtocounter{figure}{-1}

\begin{figure*}[t]
\begin{centering}
  \includegraphics[scale=0.30,trim= 20 20 20 0,clip=false]{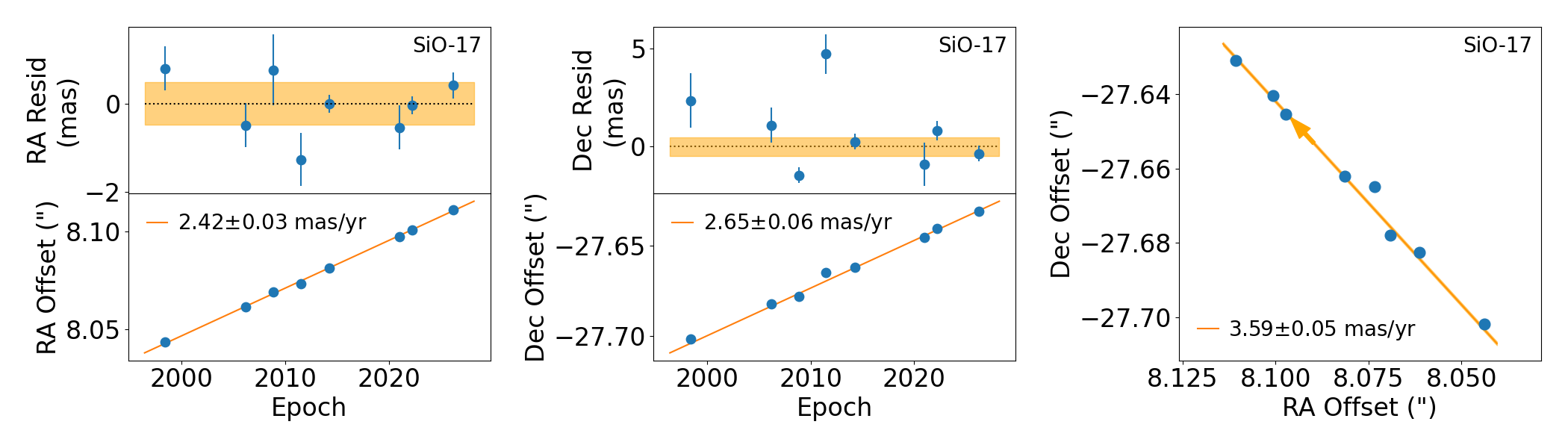}
  \includegraphics[scale=0.30,trim= 20 20 20 0,clip=false]{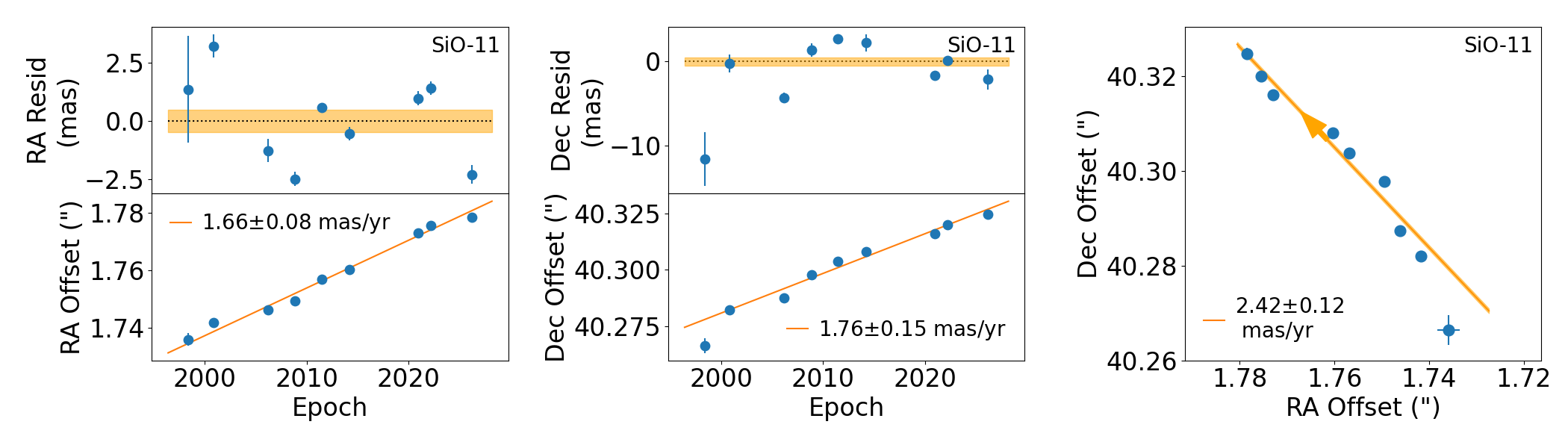}
  \includegraphics[scale=0.30,trim= 20 20 20 0,clip=false]{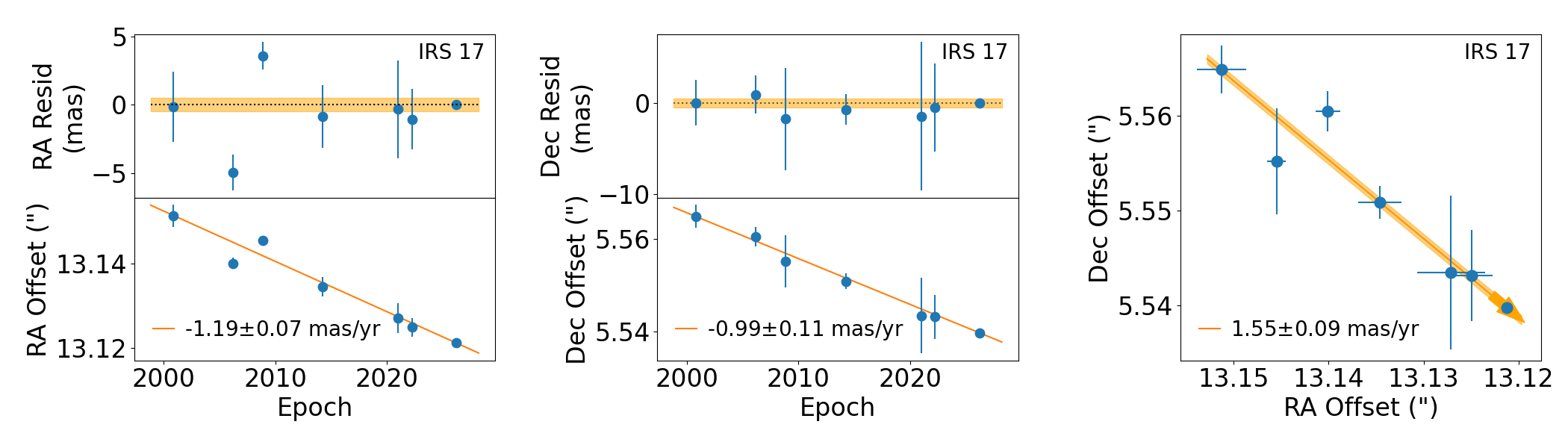}
  \includegraphics[scale=0.30,trim= 20 20 20 0,clip=false]{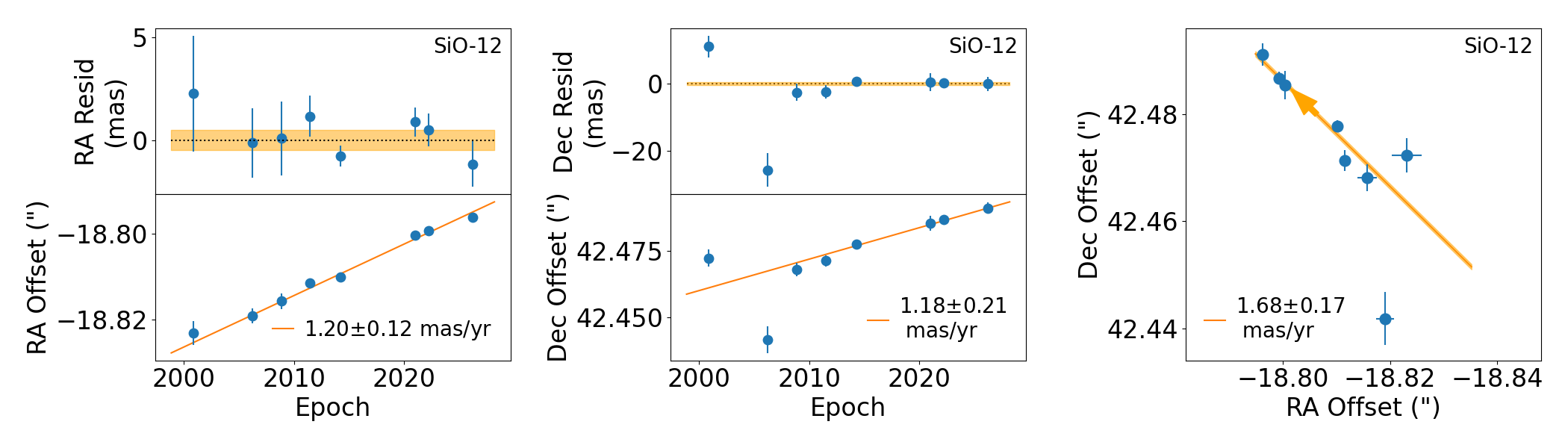}
  \includegraphics[scale=0.30,trim= 20 20 20 0,clip=false,]{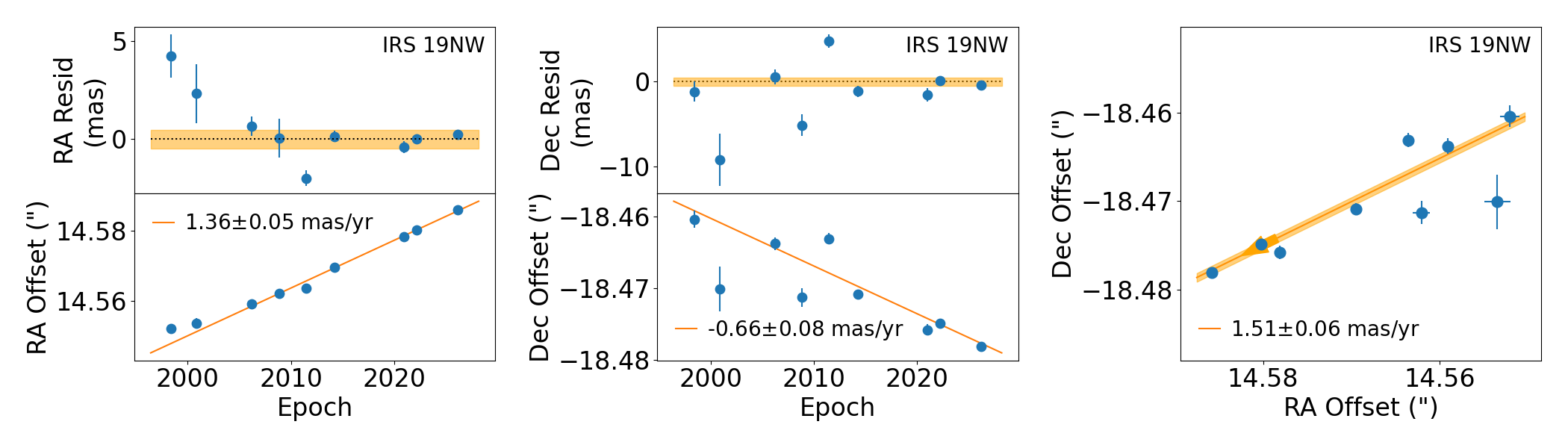}
  \caption{{\it continued}}
\end{centering}
\end{figure*}

\clearpage

\startlongtable
\begin{deluxetable*}{lrRR}
  \tablecaption{\label{tab:time_series} SiO Maser Coordinate Time Series}
  \tablehead{ \colhead{Name} & \colhead{Epoch} & \colhead{RA Offset\tablenotemark{a}} & \colhead{Dec Offset} \\
    &  & \colhead{ (arcsec)} & \colhead{(arcsec)}  } 
  \startdata
IRS 9  & 1998.410  &  5.6502 $\pm$  0.0007 &  $-$6.3503 $\pm$ 0.0013\\
          & 2000.850  &  5.6590 $\pm$  0.0011 &  $-$6.3461 $\pm$ 0.0017\\
          & 2006.200  &  5.6736 $\pm$  0.0004 &  $-$6.3341 $\pm$ 0.0009\\
          & 2008.860  &  5.6819 $\pm$  0.0007 &  $-$6.3285 $\pm$ 0.0009\\
          & 2011.470  &  5.6891 $\pm$  0.0009 &  $-$6.3205 $\pm$ 0.0008\\
          & 2014.249  &  5.6988 $\pm$  0.0002 &  $-$6.3152 $\pm$ 0.0006\\
          &  2020.988 &  5.7189 $\pm$  0.0004 &  $-$6.2998 $\pm$ 0.0009\\
          &  2022.227 &  5.7231 $\pm$  0.0001 &  $-$6.2964 $\pm$ 0.0003\\
          &  2026.208 &  5.7350 $\pm$  0.0002 &  $-$6.2874 $\pm$ 0.0006\\
  \hline
IRS 7\tablenotemark{b}  &   1998.410 &  0.0375  $\pm$  0.0043 &  5.5493  $\pm$  0.0014\\
&  2000.850 &  0.0344 $\pm$  0.0016 &  5.5418 $\pm$  0.0030\\
&  2006.200 &  0.0326  $\pm$  0.0007 &  5.5232 $\pm$   0.0013\\
&  2008.860 &  0.0345  $\pm$  0.0006 &  5.5134  $\pm$  0.0009\\
&  2011.470 &  0.0331 $\pm$  0.0007 &  5.4992 $\pm$   0.0014\\
&  2014.249 &  0.0323  $\pm$  0.0007 &  5.4872  $\pm$  0.0015\\
&  2020.988 &  0.0351  $\pm$  0.0010 &  5.4569 $\pm$   0.0034\\
&  2022.227 &  0.0331  $\pm$  0.0004 &  5.4492  $\pm$  0.0010\\
&  2026.208 &  0.0322 $\pm$  0.0004  &  5.4332  $\pm$  0.0007\\
\hline
SiO-14 &   1998.410 & $-$7.6631 $\pm$   0.0012  &$-$28.4521 $\pm$  0.0020\\
&  2000.850 & $-$7.6613  $\pm$  0.0005 & $-$28.4546 $\pm$  0.0008\\
&   2006.200 & $-$7.6478  $\pm$  0.0005&  $-$28.4557 $\pm$  0.0009\\
&   2008.860 & $-$7.6430  $\pm$  0.0011&  $-$28.4602 $\pm$  0.0015\\
&   2011.470 & $-$7.6366 $\pm$   0.0006&  $-$28.4594 $\pm$  0.0008\\
&   2014.249 & $-$7.6313 $\pm$   0.0010&  $-$28.4649 $\pm$  0.0006\\
&   2020.988 & $-$7.6176 $\pm$   0.0006&  $-$28.4720 $\pm$  0.0009\\
&   2022.227 &  $-$7.6145 $\pm$   0.0002&  $-$28.4706 $\pm$ 0.0003\\
&   2026.208 &  $-$7.6056 $\pm$  0.0001 &  $-$28.4749 $\pm$ 0.0003\\
\hline
IRS 12N &    1996.413 & $-$3.2519 $\pm$   0.0005 & $-$6.8814 $\pm$   0.0005\\
&  1998.410 & $-$3.2536  $\pm$  0.0005 & $-$6.8876  $\pm$  0.0006\\
&  2000.850 & $-$3.2559  $\pm$  0.0009 & $-$6.8939  $\pm$  0.0012\\
&  2006.200 & $-$3.2623  $\pm$  0.0009 & $-$6.9066  $\pm$  0.0019\\
&  2008.860 & $-$3.2648  $\pm$  0.0005 & $-$6.9144  $\pm$  0.0006\\
&  2011.470 & $-$3.2682  $\pm$  0.0005 & $-$6.9234  $\pm$  0.0009\\
&  2014.249 & $-$3.2713   $\pm$ 0.0002 & $-$6.9317  $\pm$  0.0005\\
&  2020.988 & $-$3.2792  $\pm$  0.0002 & $-$6.9498  $\pm$  0.0003\\
&  2022.227 & $-$3.2803  $\pm$  0.0001 & $-$6.9537  $\pm$  0.0001\\
&  2026.208 & $-$3.2843  $\pm$  0.0001 & $-$6.9646  $\pm$  0.0002\\
\hline
IRS 28   &   1998.410 & 10.4700  \pm  0.0030 & $-$5.7876 \pm   0.0050\\
&  2000.850 & 10.4698 \pm   0.0010 & $-$5.7964 \pm   0.0024\\
&  2006.200 & 10.4798 \pm   0.0007 & $-$5.8249  \pm  0.0010\\
&  2008.860 & 10.4843 \pm   0.0005 & $-$5.8393 \pm   0.0006\\
&  2011.470 & 10.4861 \pm   0.0005 & $-$5.8531 \pm   0.0010\\
&  2014.249 & 10.4918  \pm  0.0003 & $-$5.8686  \pm  0.0005\\
&  2020.988 & 10.5013  \pm  0.0014 & $-$5.9079 \pm   0.0021\\
&  2022.227 & 10.5032  \pm  0.0003 & $-$5.9119 \pm   0.0006\\
&  2026.208 & 10.5086 \pm  0.0005  & $-$5.9339 \pm   0.0009\\
\hline
SiO-15   &     2000.850 & $-$12.4266 \pm   0.0023 & $-$11.0796 \pm   0.0054\\
&  2006.200 & $-$12.4372 \pm   0.0012 & $-$11.0742  \pm  0.0015\\
&  2008.860 & $-$12.4498  \pm  0.0019 & $-$11.0744  \pm  0.0011\\
&  2014.240 & $-$12.4597 \pm   0.0004 & $-$11.0693  \pm  0.0007\\
&  2020.980 & $-$12.4770 \pm  0.0007 & $-$11.0658  \pm  0.0018\\
&  2022.220&  $-$12.4801 \pm  0.0005 & $-$11.0630 \pm  0.0005\\
&  2026.200 & $-$12.4893 \pm  0.0002 & $-$11.0611 \pm  0.0007\\
\hline
IRS 10EE  &     1996.413 &  7.6837  \pm  0.0005 &  4.2194 \pm   0.0005\\
&  1998.410 &  7.6831  \pm  0.0005 &  4.2159 \pm   0.0005\\
&  2000.850 &  7.6852  \pm  0.0005 &  4.2099  \pm  0.0005\\
&  2006.200 &  7.6832  \pm  0.0005 &  4.1986 \pm   0.0005\\
&  2008.860 &  7.6844  \pm  0.0005 &  4.1938 \pm   0.0005\\
&  2011.470 &  7.6844  \pm  0.0006 &  4.1906 \pm   0.0006\\
&  2014.249 &  7.6840  \pm  0.0003 &  4.1839 \pm   0.0004\\
&  2020.988 &  7.6851  \pm  0.0001 &  4.1697  \pm  0.0002\\
&  2022.227 &  7.6849  \pm  0.0001 &  4.1676 \pm   0.0002\\
&  2026.208 &  7.6851  \pm  0.0001 &  4.1590 \pm   0.0001\\
\hline
IRS 15NE  &     1996.413 &  1.2302 \pm   0.0005 & 11.3315 \pm   0.0005\\
&  1998.410 &  1.2264 \pm   0.0004 & 11.3197 \pm   0.0006\\
&  2000.850 &  1.2233 \pm   0.0011 & 11.3031  \pm  0.0025\\
&  2006.200 &  1.2111 \pm   0.0005 & 11.2750 \pm  0.0010\\
&  2008.860 &  1.2068  \pm  0.0006 & 11.2602 \pm   0.0007\\
&  2011.470 &  1.2023 \pm   0.0005 & 11.2448 \pm   0.0005\\
&  2014.249 &  1.1969  \pm  0.0005 & 11.2284 \pm  0.0012\\
&  2020.988 &  1.1834  \pm  0.0003 & 11.1892 \pm   0.0017\\
&  2022.227 &  1.1814 \pm   0.0003 & 11.1807 \pm  0.0004\\
&  2026.208 &  1.1733 \pm   0.0005 & 11.1581 \pm  0.0009\\
\hline
SiO-16   &    1998.410  &$-$26.4209 \pm   0.0020  &$-$34.4520  \pm  0.0027\\
&  2000.850  &$-$26.4249 \pm  0.0006  &$-$34.4425  \pm  0.0011\\
&  2006.200  &$-$26.4165 \pm  0.0006  &$-$34.4518 \pm  0.0013\\
&  2008.860  &$-$26.4214 \pm  0.0010  &$-$34.4482 \pm  0.0025\\
&  2011.470  &$-$26.4164 \pm   0.0008  &$-$34.4572 \pm   0.0007\\
&  2014.249  &$-$26.4186 \pm  0.0005  &$-$34.4666 \pm  0.0010\\
&  2022.227 & $-$26.4177  \pm  0.0002  &$-$34.4800  \pm  0.0004\\
&  2026.208 & $-$26.4166  \pm  0.0012  & $-$34.4894 \pm  0.0020\\
\hline
SiO-6    &     1998.410 & 35.2176 \pm   0.0029 & 30.6541  \pm  0.0050\\
&  2000.850 & 35.2360 \pm   0.0006 & 30.6600  \pm  0.0010\\
&  2006.200 & 35.2415 \pm   0.0010 & 30.6670 \pm   0.0028\\
 & 2008.860 & 35.2521 \pm  0.0009 & 30.6796  \pm  0.0049\\
 & 2011.470 & 35.2643 \pm   0.0020 & 30.6899 \pm   0.0040\\
&  2014.249 & 35.2687 \pm   0.0018 & 30.6997 \pm   0.0019\\
 & 2020.988 & 35.2849 \pm   0.0005 & 30.7034 \pm   0.0016\\
&  2022.227 & 35.2908 \pm   0.0008 & 30.7116 \pm   0.0019\\
&  2026.208 & 35.2991 \pm   0.0006 & 30.7165 \pm  0.0015\\
\hline
SiO-17   &    1998.410 &  8.0438 \pm   0.0005  & $-$27.7019 \pm  0.0014\\
&  2006.200 &  8.0614 \pm   0.0005  &$-$27.6825 \pm  0.0009\\
&  2008.860 &  8.0691 \pm   0.0008  &$-$27.6780 \pm  0.0004\\
&  2011.470 &  8.0734 \pm   0.0006  &$-$27.6649 \pm  0.0010\\
& 2014.249 &  8.0814  \pm  0.0002 & $-$27.6620 \pm  0.0004\\
&  2020.988 &  8.0972 \pm   0.0005 & $-$27.6453 \pm  0.0011\\
&  2022.227 &  8.1007 \pm   0.0002 & $-$27.6403 \pm  0.0005\\
&  2026.208 &  8.1108 \pm   0.0003 & $-$27.6309 \pm  0.0004\\
\hline
SiO-11   &    1998.410 &  1.7359 \pm   0.0023 & 40.2664 \pm   0.0032\\
&  2000.850 &  1.7418 \pm   0.0005 & 40.2821 \pm   0.0011\\
&  2006.200 &  1.7462  \pm  0.0005 & 40.2874 \pm   0.0006\\
&  2008.860 &  1.7494  \pm  0.0003 & 40.2978  \pm  0.0008\\
&  2011.470 &  1.7568  \pm  0.0002 & 40.3037 \pm   0.0004\\
&  2014.249 &  1.7603  \pm  0.0003 & 40.3081 \pm   0.0010\\
&  2020.988 &  1.7730 \pm   0.0003 & 40.3161 \pm   0.0005\\
&  2022.227 &  1.7755 \pm   0.0003 & 40.3200 \pm   0.0005\\
&  2026.208 &  1.7784 \pm   0.0004 & 40.3248 \pm   0.0012\\
\hline
IRS 17    &   2000.850 & 13.1513  \pm  0.0026 &  5.5649  \pm  0.0025\\
&  2006.200 & 13.1401  \pm  0.0013 &  5.5605 \pm   0.0021\\
&  2008.860 & 13.1455 \pm   0.0010 &  5.5552  \pm  0.0056\\
&  2014.249 & 13.1346 \pm   0.0023 &  5.5509  \pm  0.0017\\
&  2020.988 & 13.1271 \pm  0.0036 &  5.5435  \pm  0.0081\\
&  2022.227 & 13.1249  \pm  0.0022 &  5.5432 \pm   0.0048\\
&  2026.208 & 13.1212  \pm  0.0002 &  5.5398 \pm   0.0004\\
\hline
SiO-12   &    2000.850 & $-$18.8231  \pm  0.0028 & 42.4723  \pm  0.0032\\
&  2006.200 & $-$18.8191 \pm   0.0017 & 42.4418  \pm  0.0050\\
&  2008.860 & $-$18.8157  \pm  0.0018 & 42.4681  \pm  0.0025\\
&  2011.470  &$-$18.8115 \pm   0.0010 & 42.4713 \pm   0.0020\\
&  2014.249  &$-$18.8101  \pm  0.0005 & 42.4777 \pm   0.0011\\
&  2020.988  &$-$18.8004  \pm  0.0007 & 42.4854  \pm  0.0027\\
&  2022.227 & $-$18.7993  \pm  0.0008 & 42.4867  \pm  0.0012\\
&  2026.208 & $-$18.7962  \pm  0.0011 & 42.4911  \pm  0.0021\\
\hline
IRS 19NW  &   1998.410 & 14.5521  \pm  0.0011 & $-$18.4604\pm  0.0012\\
&  2000.850 &14.5535 \pm 0.0015 & $-$18.4701  \pm  0.0031\\
&  2006.200 & 14.5591  \pm  0.0005 & $-$18.4638 \pm  0.0009\\
&  2008.860 &  14.5621 \pm  0.0010 & $-$18.4713  \pm  0.0013\\
&  2011.470 & 14.5636 \pm   0.0004 & $-$18.4631  \pm  0.0008\\
&  2014.249 & 14.5695  \pm  0.0003& $-$18.4709 \pm  0.0007\\
&  2020.988 & 14.5781 \pm  0.0003 & $-$18.4758 \pm   0.0008\\
&  2022.227 & 14.5802  \pm  0.0001 & $-$18.4749 \pm   0.0002\\
&  2026.208 & 14.5858  \pm  0.0002 & $-$18.4781 \pm   0.0004\\
\enddata
\tablecomments{Coordinate offsets are with respect to the Sgr~A* radio centroid at each epoch.  Unlike Table~\ref{tab:time_series}, the relative differential offsets described in Section \ref{sec:methods} have \emph{not} been subtracted here; they have instead been restored (added back) into the coordinate offsets, so the values listed are the uncorrected positions.}
\tablenotetext{a}{This offset is corrected for Declination:  it is $\Delta {\rm RA} \cos({\rm Dec})$.}
\tablenotetext{b}{The first two epochs were not used for the IRS~7 astrometry but are included here for posterity.}
\end{deluxetable*}

\section{Proper Motion Measurement Methods}\label{sec:pms}

\citet{darling2023} describes in detail several methods for obtaining proper motions from astrometric time series and finds that these generally do not
provide significantly different outcomes from simple variance-weighted least-squares fits.  In this work, we re-examined the methods and again find that
the conventional method remains adequate.  Nonetheless, sample comparisons are presented here.  The fitting methods are:
(1) a variance-weighted least-squares;
(2) a ``conservative formulation'' of uncertainties that allows for large outliers in the data \citep{sivia2006, darling2018};
(3) a ``good-and-bad data'' model \citep{box1968,sivia2006}; and
(4) a variance-weighted least-squares with expanded uncertainties as a free parameter \citep{emcee}.

\begin{figure*}[ht!]
\begin{centering}
  \includegraphics[scale=0.5,trim= 0 44 20 0,clip=false]{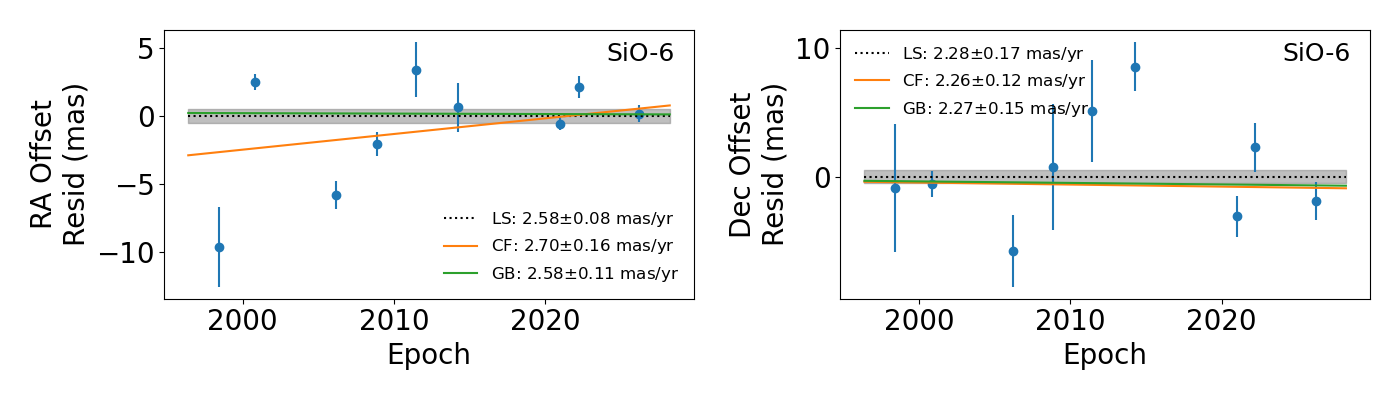}
  \includegraphics[scale=0.5,trim= 0 20 20 0,clip=true]{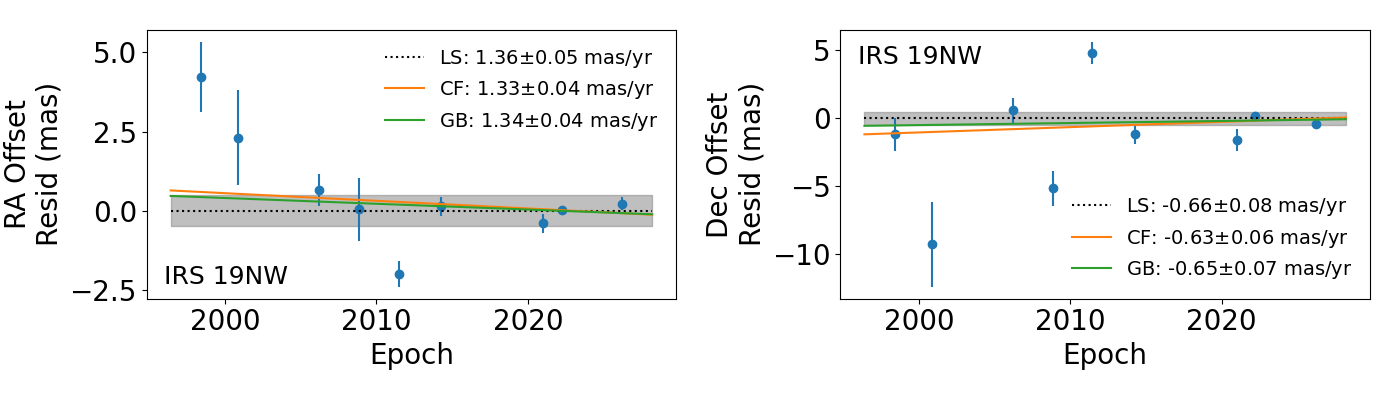}
  \caption{Example residual proper motion time series showing the three fitting methods (LS = least-squares; CF = conservative formulation; GB = good-and-bad data).
  The residual is with respect to the weighted least-squares fit, and the shaded regions indicate $\pm4$~AU.
  Top:    SiO-6 shows differing proper motion solutions in RA for all three methods.
  Bottom:  IRS 19NW, showing proper motion solutions for the conservative formulation and good-and-bad methods that differ from
  the canonical least-squares fit.
  For most of the masers, the two less conventional fitting methods do not improve upon or differ significantly from 
  the least-squares method.}
\label{fig:PMresids}
\end{centering}
\end{figure*}

\begin{figure*}[ht!]
\begin{centering}
  \includegraphics[scale=0.5,trim= 0 44 20 0,clip=false]{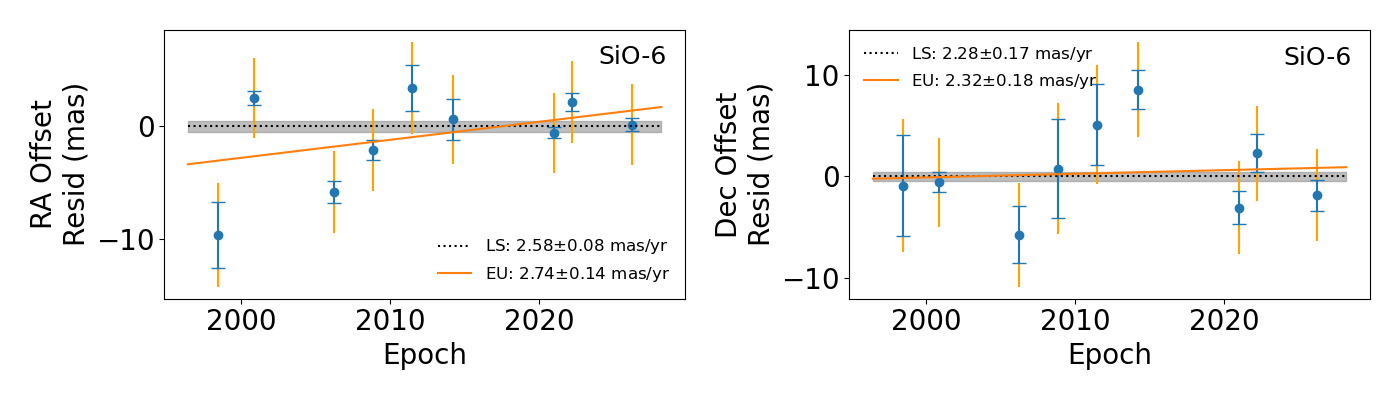}
  \includegraphics[scale=0.5,trim= 0 20 20 0,clip=true]{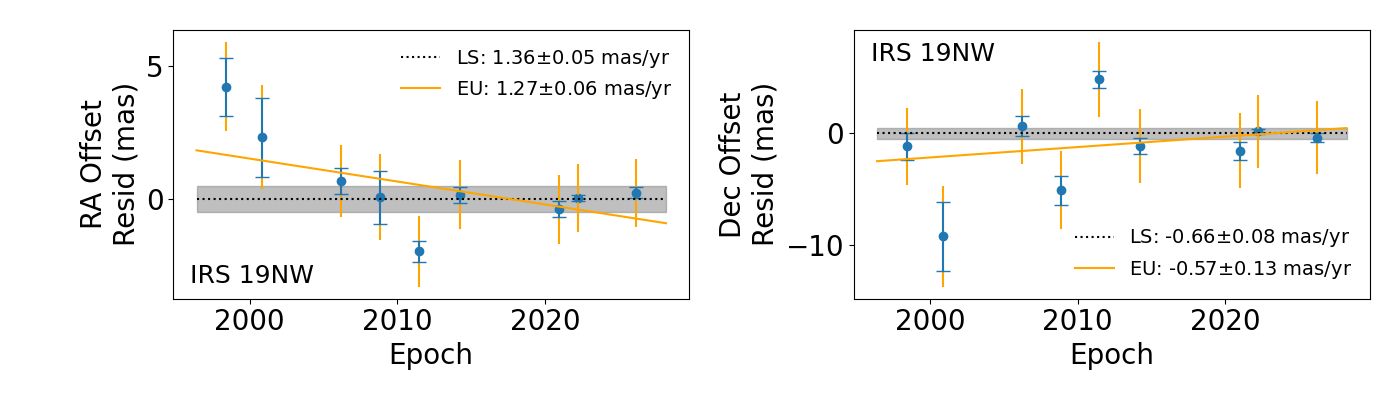}
  \caption{Example residual proper motion time series showing the result of 
    expanded uncertainties (EU) that are added in quadrature to the measurement uncertainties.
    The residual is with respect to the weighted least-squares (LS) fit, and the shaded regions indicate $\pm4$~AU.
    The EU proper motions do not significantly differ from the LS proper motions ($\lesssim 1\sigma$).
    Top:  SiO-6 has expanded uncertainties of 3.5 and 4.3 mas, respectively.
    Bottom: The additional uncertainties are 1.3 mas in RA and 3.2 mas in Dec for IRS 19NW.
    }
\label{fig:PMresidsEU}
\end{centering}
\end{figure*}

Figure \ref{fig:PMresids} shows examples of the proper motion fit residuals for the first three methods
for two masers that show scatter inconsistent with their astrometric uncertainties, IRS~19NW and SiO-6.  
Even for these extreme cases, the proper motions obtained from the fitting methods (2) and (3) are formally consistent with the
simple least-squares fit.  Figure \ref{fig:PMresidsEU} shows proper motion fit residuals for methods (1) and (4) for the same masers.  In this
case, expanded uncertainties in the astrometry are favored, but the proper motions do not significantly differ ($\lesssim 1\sigma$)
in part due to larger uncertainties in the method (4) proper motions.  As seen in \citep{darling2023}, the expanded uncertainties are larger
than the expected size of the maser-emitting regions, so the astrometric scatter cannot be explained by changing structure of maser-emitting regions.

For the entire maser ensemble, expanded uncertainties are favored in roughly half of the sample, which is consistent with the distribution of $\chi^2_\nu$ values in Table \ref{tab:results2026}.   The proper motions obtained from expanding the uncertainties are never significantly different from the least-squares fits to the astrometric time series, and the reference frame stability changes by only $\sim$10\% in each coordinate.  



\end{document}